\documentclass[twocolumn, switch]{article} 

\usepackage{preprint}

\usepackage{cite}
\usepackage{amsmath,amssymb,amsfonts}
\usepackage{graphicx}
\usepackage{textcomp}
\PassOptionsToClass{usenames,dvipsnames,table,xcdraw}{xcolor}
\usepackage{xcolor}

\usepackage[ruled,boxed]{algorithm2e}
\usepackage{algpseudocode}
\usepackage{adjustbox}
\usepackage{caption}
\usepackage{subcaption}
\usepackage{stfloats}
\usepackage{multirow}
\usepackage{tabularx}
\usepackage{listings}
\usepackage{comment}
\usepackage{soul}
\usepackage[utf8]{inputenc}	
\usepackage[T1]{fontenc}	
\usepackage[colorlinks = true,
            linkcolor = purple,
            urlcolor  = blue,
            citecolor = cyan,
            anchorcolor = black]{hyperref}	
\usepackage{booktabs} 		
\usepackage{nicefrac}		
\usepackage{microtype}		
\usepackage{lineno}		

\newenvironment{algomethod}[1][]
  {\renewcommand{\algorithmcfname}{Method}%
   \begin{algorithm}[#1]
   \long\def\@caption##1[##2]##3{%
     \par
     \begingroup\@parboxrestore
     \if@minipage\@setminipage\fi
     \normalsize \@makecaption{\AlCapSty{\AlCapFnt\algorithmcfname}}{\ignorespaces ##3}%
     \par\endgroup
   }}
  {\end{algorithm}}
\SetKwFor{For}{for (}{) $\lbrace$}{$\rbrace$}

\newcommand{\paragraphbe}[1]{\par{\textbf{#1.}}}

\def\BibTeX{{\rm B\kern-.05em{\sc i\kern-.025em b}\kern-.08em
    T\kern-.1667em\lower.7ex\hbox{E}\kern-.125emX}}

\usepackage{titlesec}
\titlespacing\section{0pt}{12pt plus 3pt minus 3pt}{1pt plus 1pt minus 1pt}
\titlespacing\subsection{0pt}{10pt plus 3pt minus 3pt}{1pt plus 1pt minus 1pt}
\titlespacing\subsubsection{0pt}{8pt plus 3pt minus 3pt}{1pt plus 1pt minus 1pt}

\title{A Depolarizing Noise-aware Transpiler for Optimal Amplitude Amplification}

\author{
  Debashis Ganguly \\
  Independent Researcher \\
  \texttt{DebashisGanguly@gmail.com} \\
   \And
 Wonsun Ahn \\
  Department of Computer Science\\
  University of Pittsburgh \\
  \texttt{wonsun.ahn@gmail.com} \\
}

\begin{document}

\twocolumn[ 
  \begin{@twocolumnfalse} 
  
\maketitle

\begin{abstract}
Amplitude amplification provides a quadratic speed-up for an array of quantum algorithms when run on a quantum machine perfectly isolated from its environment. However, the advantage is substantially diminished as the NISQ-era quantum machines lack the large number of qubits necessary to provide error correction. Noise in the computation grows with the number of gate counts in the circuit with each iteration of amplitude amplification. After a certain number of amplifications, the loss in accuracy from the gate noise starts to overshadow the gain in accuracy due to amplification, forming an inflection point. Beyond this point, accuracy continues to deteriorate until the machine reaches a maximally mixed state where the result is uniformly random. Hence, quantum transpilers should take the noise parameters of the underlying quantum machine into consideration such that the circuit can be optimized to attain the maximal accuracy possible for that machine. In this work, we propose an extension to the transpiler that predicts the accuracy of the result at every amplification with high fidelity by applying pure Bayesian analysis to individual gate noise rates.  Using this information, it finds the inflection point and optimizes the circuit by halting amplification at that point.  The prediction is made without needing to execute the circuit either on a quantum simulator or an actual quantum machine.

\end{abstract}
\vspace{0.35cm}

  \end{@twocolumnfalse} 
] 

\section{Introduction}
\label{sec:introduction}

Lov Grover's quantum search algorithm~\cite{grover1996fast} has initially been proposed for unstructured search problems, i.e., for finding a marked element in a unstructured database with high probability. However, today, its uses extend beyond that; as it can be used as a general subroutine to obtain quadratic run time improvements for a variety of other algorithms~\cite{kwon2021quantum,rajagopal2021quantum,franco2009quantum,elias2021enhanced,koch2022gaussian,bera2018amplitude,daoyi2007reinforcement}. Although Grover’s algorithm does not solve NP–complete problems in polynomial time, the wide range of applications, that use it as a subroutine more than compensates for this.
 
 Amplitude amplification~\cite{brassard1997exact} is a generalization of the basic idea behind Grover's algorithm.  At a high level, amplitude amplification works in an iterative manner where, at each iteration, the probability of the winning outcome is amplified.  The probability approaches one after $O(\sqrt{N})$ iterations, where $N$ is the size of the search space, leading to a quadratic speedup over classical algorithms.
 Section~\ref{subsec:amp_amplification} covers the mathematical foundation of amplitude amplification in depth.

Today's quantum processors are ``noisy'' and may lose their quantum state due to quantum decoherence over time. Moreover, they do not have enough qubits to perform error correction on the noise.  These are the two defining characteristics of noisy intermediate-scale quantum (NISQ) machines. Amplitude amplification involves repeating the oracle and diffuser circuits in Figure~\ref{fig:CircuitAmplitudeAmplification} iteratively.  Hence. the total number of gates in the circuit grows linearly with each iteration, and with it, gate noise.  After a certain number of iterations, loss in accuracy due to gate noise trumps the gain in accuracy from amplitude amplification.

Most quantum machines currently are part of a cloud infrastructure where many machines with varying qubit numbers and noise parameters are assembled as a shared resource.  This is done to amortize the cost of building and maintaining a quantum infrastructure over many users.  When users submit jobs to the cloud, it is the responsibility of the job scheduler to find an appropriate machine to run on, or reject the job if it is not suitable for any machine. As of now, the only hard criterion that the scheduler can use is the number of qubits. Hence, it considers any machine with the requisite number of qubits to be equally suitable. Machines do expose some noise parameters to the scheduler such as gate noise rates for each physical gate in the machine and thermal relaxation times; but there is no framework which can help predict how those parameters will affect the accuracy of the result.

We propose an extension to the transpiler, which converts a given quantum algorithm to a quantum circuit for the given machine. The purpose of the extension is to 1) given the transpiled circuit, \emph{provide a prediction} on the accuracy of the final outcome by analyzing the circuit and applying the noise parameters to each gate, and 2) \emph{optimize the transpiled circuit} so that it reaches the maximal accuracy that is possible on that machine, based on that prediction. Armed with this information, the scheduler now can assign a job to the machine that is most suitable for the job and also be guaranteed that the job will be executed in the most optimal way on that machine.

Our transpiler extension uses a Bayesian probabilistic model to compute the probability of noise seeping into an amplification iteration, analyzing the circuit using noise parameters of single-qubit and two-qubit quantum gates advertised by the quantum machine. It makes this estimation for every iteration of an amplitude amplification loop. The accuracy measured at every iteration initially increases due to amplitude amplification but often
displays an ``inflection point'' beyond which further amplification cannot surmount the degrading effects of noise.  If an inflection point exists, the transpiler optimizes this circuit by truncating the circuit at that point so that no more iterations are performed.

Now, it is worth mentioning that the probability of noise-free execution is not equivalent to the probability of the outcome being accurate. That is because even noisy outcomes have the potential to produce the winning result when measured. This is true for noisy quantum systems as it is true for noisy classical systems.  However, we will show the probability of noise-free execution is closely correlated to the accuracy of the result, such that making a prediction based on the former still results in a near optimal circuit.

In summary, we make the following contributions:

\begin{itemize}
    \item We make the empirical observation that there sometimes exists an inflection point beyond which further amplification iterations result in losses in accuracy.
    \item We propose novel intrinsics in the OpenQASM language that delineates iterations in an iterative quantum algorithm such as amplitude amplification and also conveys the amplification achieved in each iteration.
    \item We implement an extension to the circuit transpiler that utilizes Bayesian probabilistic model to estimate the depolarizing noise generated by the gates in the iteration delineated by the intrinsics, in order to predict the inflection point. 
	\item We demonstrate that the estimated probability of the winning outcome(s) being noise-free follows closely the probability of the result being winning outcome(s), by actual measurements on quantum circuit. As a result, we show that our prediction of the inflection point is highly accurate.
\end{itemize}

\section{Background}
\label{sec:background}

First, we will cover the necessary background on amplitude amplification to give the readers a comprehensive perspective on the iterative execution of quantum circuit to find a solution with higher probability. Then, this section describes depolarizing channel, one of the most common noise models in quantum computing considered by real quantum resources.

\subsection{Amplitude Amplification}
\label{subsec:amp_amplification}

Grover's quantum search algorithm uses a procedure called amplitude amplification, which is how a quantum computer significantly enhances the probability of a solution state. This procedure amplifies or stretches out the amplitude of the marked item(s), which shrinks the amplitude of the other items, such that measuring the final state will return the right item(s) with near-certainty. This section explains the basic steps invoked by Grover's quantum search algorithm as part of amplitude amplification, highlighting the geometrical interpretation in terms of two reflections, which generate a rotation in a two-dimensional plane. 

Grover's algorithm is an example of an unstructured quantum search where we wish to locate one item with a unique property in a large list of $N$ items. At the beginning, we have no idea where the marked item is. Therefore, any guess of its location is as good as any other, which can be expressed in terms of a uniform superposition:  $\lvert s \rangle = \frac{1}{\sqrt{N}}\sum_{x=0}^{N-1}\lvert x \rangle$, where the dimension of the problem is $N$. If at this point we were to measure in the standard basis $\{\lvert x \rangle\}$, this superposition would collapse, according to the fifth quantum law, to any one of the basis states with the same probability of $\frac{1}{N}=\frac{1}{2^n}$, where $n$ is the number of qubits to compose the problem. Our chances of guessing the right value is therefore $1$ in $2^n$, as could be expected. Hence, on average we would need to try about $\frac{N}{2} = 2^{n-1}$ times to guess the correct item by a classical algorithm. 

\begin{figure}[h!t]
    \centering
    \subcaptionbox{Initial superposition state\label{fig:InitialSuperpositionState}}
        {\includegraphics[width=\linewidth]{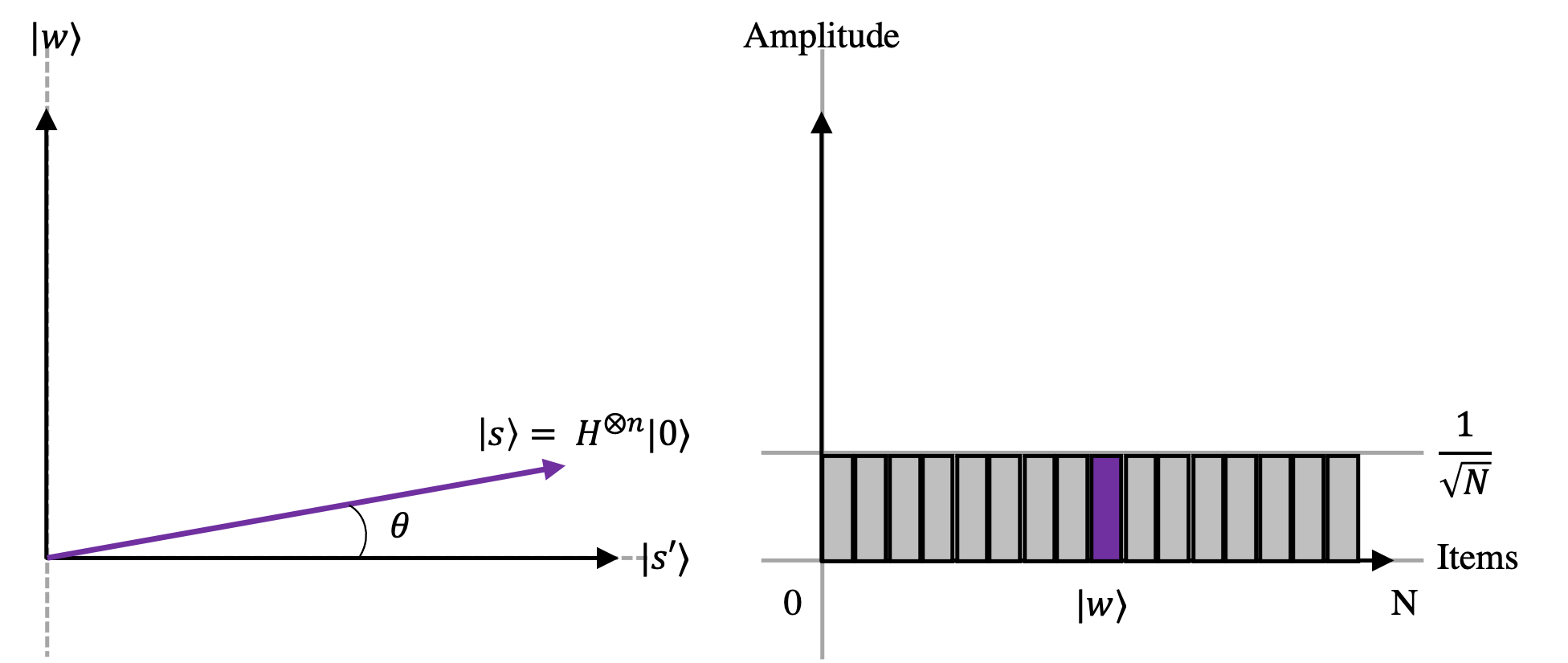}}
    \subcaptionbox{Conditional phase shift by oracle\label{fig:OracleConditionalPhaseShift}}
        {\includegraphics[width=\linewidth]{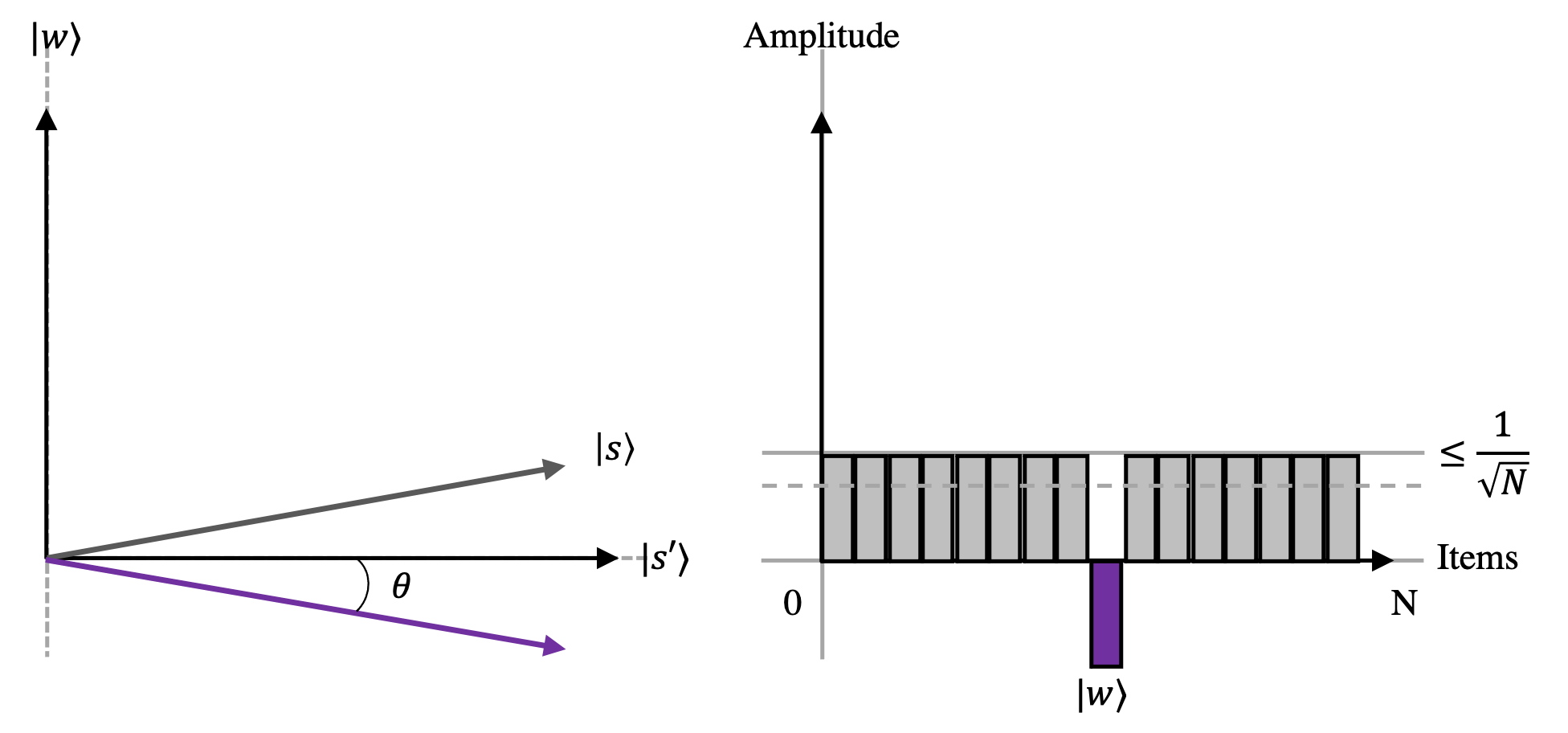}}
    \subcaptionbox{Inversion about the mean\label{fig:InversionAboutMean}}
        {\includegraphics[width=\linewidth]{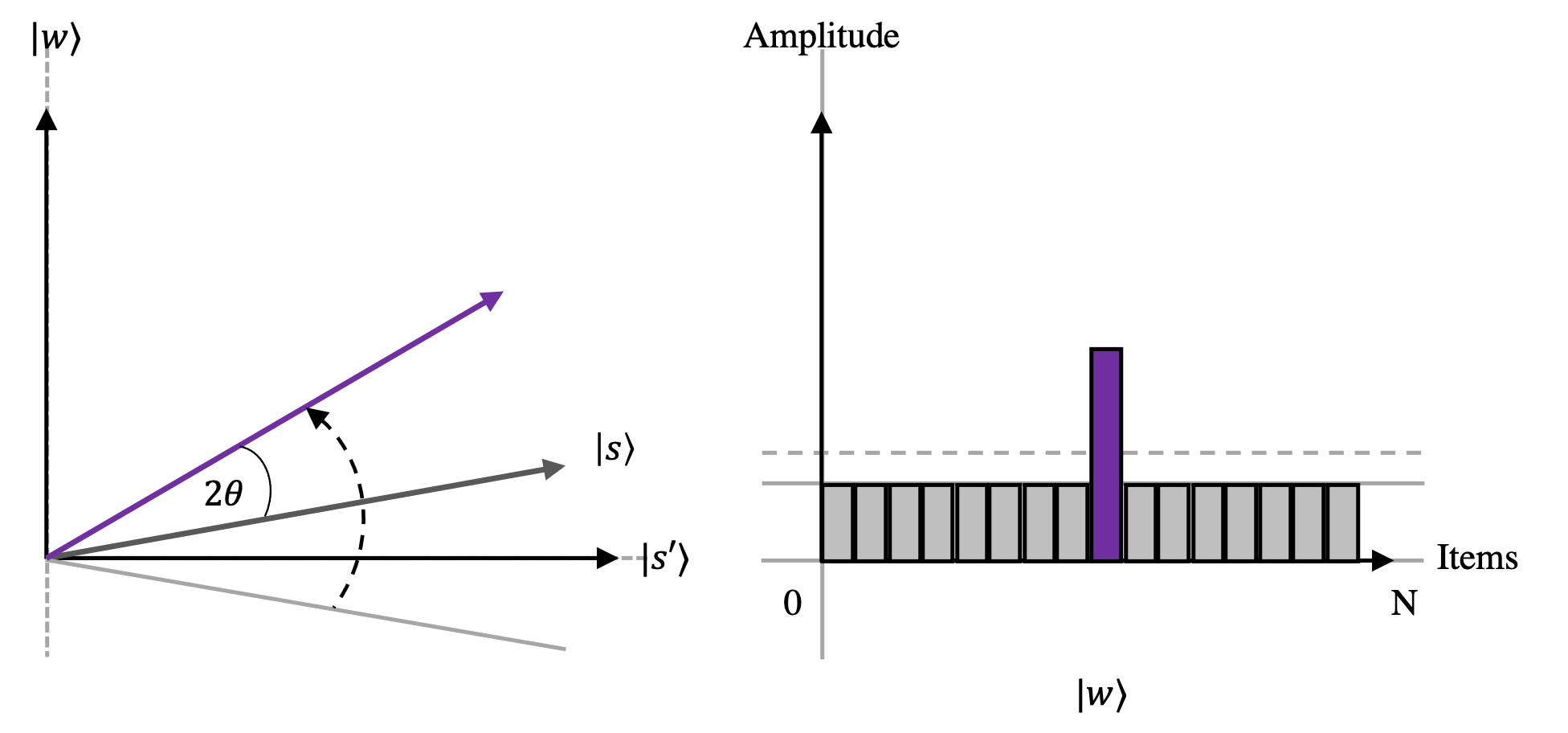}}
    \caption{Geometrical representation of amplitude amplification.}
    \label{fig:amplitude_amplification}
\end{figure}

\paragraphbe{Step 1: Initial superposition state} The amplitude amplification procedure starts out in the uniform superposition $\lvert s \rangle$, which is constructed from $\lvert s \rangle = H^{\otimes n}{\lvert0\rangle}^n$. The left graphic in Figure~\ref{fig:InitialSuperpositionState} corresponds to the two-dimensional plane spanned by perpendicular vectors $\lvert w \rangle$ and $\lvert s' \rangle$. This allows to express the initial state as  $\lvert s \rangle = \sin(\theta)\lvert w \rangle + \cos(\theta)\lvert s' \rangle$, where $\theta = \arcsin(\langle s \lvert w \rangle)$, and $\lvert w \rangle$ and $\lvert s' \rangle$ are respectively the winner and an additional state perpendicular to $\lvert w \rangle$ obtained from $\lvert s \rangle$ by removing $\lvert w \rangle$ and rescaling.The right graphic in Figure~\ref{fig:InitialSuperpositionState} denotes a bar graph of the amplitudes of the state $\lvert s \rangle$.

\paragraphbe{Step 2: Conditional phase shift by oracle} Next, the oracle reflection $U_f$ is applied to the state $\lvert s \rangle$. This oracle, described as $U_f\lvert x \rangle = (-1)^{f(x)}\lvert x \rangle$, is a diagonal matrix, where the entry that corresponds to the marked item, that we are searching for, will have a negative phase. Geometrically this corresponds to a reflection of the state $\lvert s \rangle$ about $\lvert s' \rangle$ shown in the left graphic of Figure~\ref{fig:OracleConditionalPhaseShift}. This transformation means that the amplitude in front of the state corresponding to the marked item becomes negative, which in turn means that the average amplitude, indicated by a dashed line in right graphic of Figure~\ref{fig:OracleConditionalPhaseShift} has been lowered.

\paragraphbe{Step 3: Inversion about the mean by diffuser} Next, the diffuser circuit maps the winning state, corresponding to the marked item, to $U_{s}U_{f}\lvert s \rangle$. $U_{s} = 2\lvert s \rangle \langle s \rvert - 1$ corresponds to an additional rotation about the state $\lvert s \rangle$. The two reflections by $U_{f}$ and $U_{s}$ geometrically corresponds to a rotation, shown in left graphic of Figure~\ref{fig:InversionAboutMean}, that brings the initial state $\lvert s \rangle$ closer towards the winner $\lvert w \rangle$. Since the average amplitude has been lowered by the first reflection, $U_{f}$, this transformation boosts the negative amplitude of  $\lvert w \rangle$ to roughly three times its original value, while it decreases the other amplitudes, as shown in the right graphic of Figure~\ref{fig:InversionAboutMean}.

\begin{figure}[h!t]
\begin{center}
\includegraphics[width=0.8\linewidth]{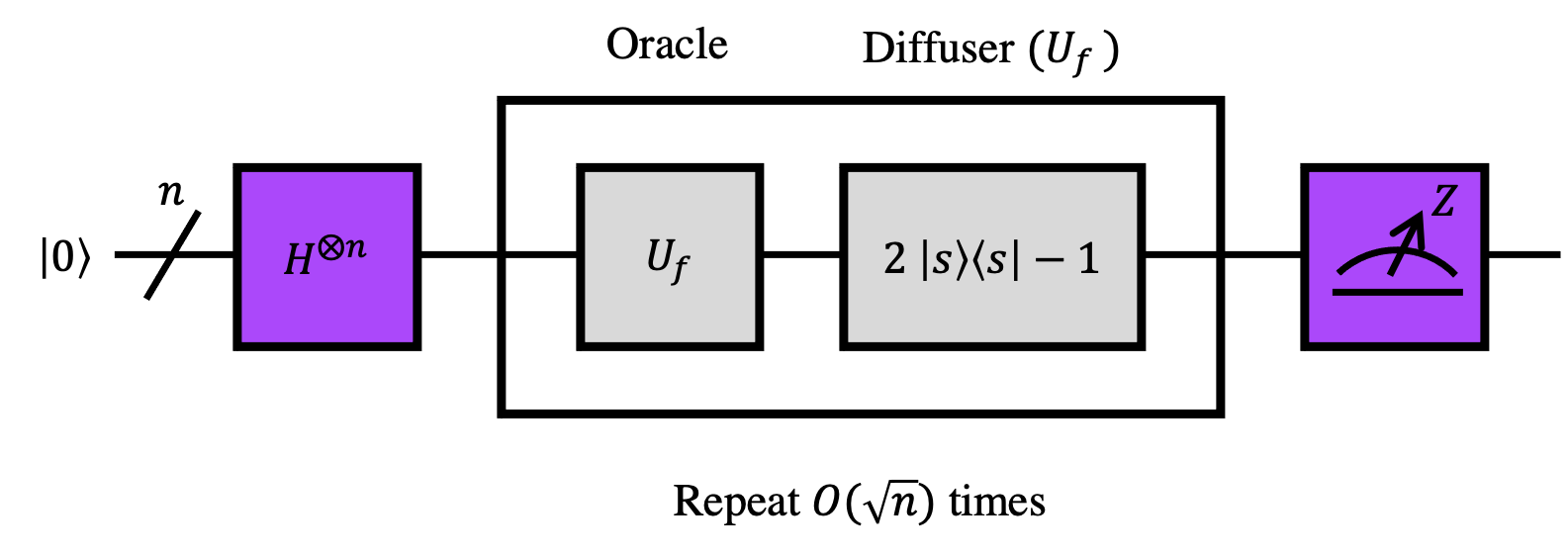}
\end{center}
\caption{Repetitive execution of quantum circuit for amplitude amplification.}
\label{fig:CircuitAmplitudeAmplification}
\end{figure}

The procedure of amplitude amplification corresponds to the combination of \textbf{Step 2} and \textbf{Step 3}, which will be repeated several times (as shown in Figure~\ref{fig:CircuitAmplitudeAmplification}) to zero in on the winner, i.e, when $\lvert s \rangle$ overlaps on $\lvert w \rangle$. It turns out the number of iterations required for this is $O(\sqrt{N})$ and thus providing a quadratic speed up over a classical counterpart.

Grover’s algorithm uses Hadamard gates, $H^{\otimes n}{\lvert0\rangle}^n$ to create the uniform superposition of all the states at the beginning of the Grover operator. If some information on the good states is available, it might be useful to not start in a uniform superposition but only initialize specific states.  Also, the diffusion operator does not reflect about the equal superposition state, but another state specified via an operator $\mathcal{A}$, where $\mathcal{Q} = \mathcal{A} \mathcal{S}_0 \mathcal{A}^\dagger \mathcal{S}_f$. This generalized version of Grover's operator ($\mathcal{Q}$) is known as Quantum Amplitude Amplification~\cite{brassard2002quantum}.

\subsection{Quantum Depolarizing Channel}
\label{subsec:quantum_depolarizing_channel}

In quantum information theory, a quantum channel is a communication channel which can transmit quantum information, as well as classical information. A quantum depolarizing channel is a model for quantum noise in quantum systems. The depolarizing channel can be viewed as a completely positive trace-preserving map, $E_{\lambda}(\rho)$, which maps a mixed state $\rho$  onto a linear combination of itself and the maximally mixed state and can be defined as

\begin{equation}
\label{eq:depolarizing_channel}
    \begin{aligned}
        E_{\lambda}(\rho)&=(1-\lambda)\rho + {\lambda}{Tr[\rho]}{\frac{I}{2^n}} \\
        \text{with } & 0 \leq \lambda \leq \frac{4^n}{4^n - 1} \\
        \text{where, } &\\
        \lambda & \text{ is the depolarizing error parameter} \\ 
        n & \text{ is the number of qubits.} \\
    \end{aligned}
\end{equation}

When $\lambda = 0$, $E_{\lambda}(\rho) = \rho$, which is the identity channel and when $\lambda = 1$, $E_{\lambda}(\rho) = \frac{I}{2^n}$, which is a completely depolarizing channel returning maximally mixed state. 

Depolarizing channel is symmetrical across all three $x$, $y$, and $z$-axes. Hence, $E_{\lambda}(\rho)$ can be interpreted as a uniform contraction of the Bloch sphere, parameterized by $\lambda$. For example, $\lambda = 0$ represents the complete  contraction of the Bloch-sphere down to the single-point $\frac{I}{2}$ given by the origin.

The additivity of Holevo information for all channels was a famous open conjecture in quantum information theory, but it is now known that this conjecture doesn't hold in general. This was proved by showing that the additivity of minimum output entropy for all channels doesn't hold~\cite{hastings2009superadditivity}, which is an equivalent conjecture. Nonetheless, the additivity of the Holevo information is shown to hold for the quantum depolarizing channel. Christopher King~\cite{king2003capacity} showed that the maximum output p-norm of the depolarizing channel is multiplicative, which implied the additivity of the minimum output entropy, which is equivalent to the additivity of the Holevo information.

\section{Motivation}
\label{sec:motivation}

This section motivates our research by presenting an empirical observation on how noise affects the outcome probability of a marked item produced by a quantum algorithm running amplitude amplification iteratively.

NISQ era quantum resources are far from perfect and riddled with noises. Users can effectively query a quantum resource to retrieve its properties such as noise parameters for different quantum gates listed by their indices or positions. There are two types of noise channels captured by these properties - (i) depolarizing channel, and (ii) thermal relaxation times. In this research, we focus on depolarizing channel alone. An important fact to note here is that the value of noise parameters vary from one position to another even for the same gate type. Table~\ref{tab:error_rate} lists the average value of quantum depolarizing noise parameters over all indices of \verb|sx|, \verb|rz|, and \verb|cx| gates present in an array of popular IBMQ quantum resources~\cite{IBMQResources}. Note the large variance not only in the number of supported qubits but also in the noise parameters, making it hard for the scheduler to optimally schedule jobs without a good predictor.

\begin{table*}[h!t]
\caption{Number of gates in the transpiled circuit}
\label{tab:gate_count}
\centering
\begin{adjustbox}{width=\textwidth,keepaspectratio}
\begin{tabular}{ll|lll|lll|lll|}
\cline{3-11}
                                            &                           & \multicolumn{3}{c|}{\textbf{rz}}                                                                                  & \multicolumn{3}{c|}{\textbf{sx}}                                                                                  & \multicolumn{3}{c|}{\textbf{cx}}                                                                                  \\ \hline
\multicolumn{1}{|l|}{\textbf{\# of qubits}} & \textbf{\# of iterations} & \multicolumn{1}{l|}{\textbf{Initial}} & \multicolumn{1}{l|}{\textbf{Inc./Iteration}} & \textbf{Total} & \multicolumn{1}{l|}{\textbf{Initial}} & \multicolumn{1}{l|}{\textbf{Inc./Iteration}} & \textbf{Total} & \multicolumn{1}{l|}{\textbf{Initial}} & \multicolumn{1}{l|}{\textbf{Inc./Iteration}} & \textbf{Total} \\ \hline
\multicolumn{1}{|l|}{\textbf{5}}            & 4                         & \multicolumn{1}{l|}{10}                      & \multicolumn{1}{l|}{106}                          & 434            & \multicolumn{1}{l|}{5}                       & \multicolumn{1}{l|}{18}                           & 75             & \multicolumn{1}{l|}{0}                       & \multicolumn{1}{l|}{80}                           & 320            \\ \hline
\multicolumn{1}{|l|}{\textbf{7}}            & 8                         & \multicolumn{1}{l|}{14}                      & \multicolumn{1}{l|}{402}                          & 3230           & \multicolumn{1}{l|}{7}                       & \multicolumn{1}{l|}{14}                           & 119            & \multicolumn{1}{l|}{0}                       & \multicolumn{1}{l|}{376}                          & 3008           \\ \hline
\multicolumn{1}{|l|}{\textbf{9}}            & 17                        & \multicolumn{1}{l|}{18}                      & \multicolumn{1}{l|}{1562}                         & 26572          & \multicolumn{1}{l|}{9}                       & \multicolumn{1}{l|}{18}                           & 315            & \multicolumn{1}{l|}{0}                       & \multicolumn{1}{l|}{1528}                         & 25976          \\ \hline
\end{tabular}
\end{adjustbox}
\end{table*}

\begin{table}[h!t]
\caption{Error parameters for depolarizing channel reported from IBM Quantum Compute Resources}
\label{tab:error_rate}
\centering
\begin{tabular}{ll|lll|}
\cline{3-5}
                                           &                       & \multicolumn{3}{l|}{\textbf{Average Error Rate}}                                  \\ \hline
\multicolumn{1}{|l|}{\textbf{System Name}} & \textbf{\# of Qubits} & \multicolumn{1}{l|}{\textbf{rz}} & \multicolumn{1}{l|}{\textbf{sx}} & \textbf{cx} \\ \hline
\multicolumn{1}{|l|}{Washington}           & 127                   & \multicolumn{1}{l|}{0}           & \multicolumn{1}{l|}{0.001}       & 0.03882     \\ \hline
\multicolumn{1}{|l|}{Brooklyn}             & 65                    & \multicolumn{1}{l|}{0}           & \multicolumn{1}{l|}{0.0005}      & 0.01279     \\ \hline
\multicolumn{1}{|l|}{Toronto}              & 27                    & \multicolumn{1}{l|}{0}           & \multicolumn{1}{l|}{0.0007}      & 0.02083     \\ \hline
\multicolumn{1}{|l|}{Guadalupe}            & 16                    & \multicolumn{1}{l|}{0}           & \multicolumn{1}{l|}{0.0004}      & 0.01079     \\ \hline
\multicolumn{1}{|l|}{Nairobi}              & 7                     & \multicolumn{1}{l|}{0}           & \multicolumn{1}{l|}{0.0003}      & 0.00788     \\ \hline
\multicolumn{1}{|l|}{Santiago}             & 5                     & \multicolumn{1}{l|}{0}           & \multicolumn{1}{l|}{0.0002}      & 0.00603     \\ \hline
\end{tabular}
\end{table}

A $Z$-gate implements rotations around the $Z$ axis which corresponds to a change in the relative phase between the $\lvert0\rangle$ and $\lvert1\rangle$ states. A $Z$ gate can be implemented by either detuning the frequency of the qubit with respect to the drive field for some finite amount of time or by composite $X$ and $Y$ gates. A $Z$ gate, implemented by adding a phase offset to the drive field for all subsequent $X$ and $Y$ gates, is essentially perfect~\cite{mckay2017efficient}. IBM Quantum Compute Resources utilize these virtual $Z$ gates to improve overall fidelity of the system (Table~\ref{tab:error_rate}).

The goal of this work is to find the optimal number of iterations for amplitude amplification in presence of noise. The noise from depolarizing channel accumulates multiplicatively over the number of quantum gates each qubit goes though. Thus, it is of paramount importance to consider the number of quantum gates present in a tanspiled circuit before predicting the optimal number of iterations. As shown in Figure~\ref{fig:CircuitAmplitudeAmplification}, quantum algorithms such as Grover's quantum search, that utilizes amplitude amplification as a subtoutine, repeats the oracle and diffuser circuit over multiple iterations to find the marked element with higher probability. As a result, starting with the initial circuit that creates an equal superposition state by applying  Hadamard gates, $H^{\otimes n}{\lvert0\rangle}^n$, with each iteration of amplitude amplification, the number of gates grows linearly. Moreover, with increase in dimensionality of the problem or the number of qubits, the number of quantum gates can increase exponentially. Table~\ref{tab:gate_count} lists the number of \verb|sx|, \verb|rz|, and \verb|cx| gates for a SAT solver using Grover's algorithm for $5$, $7$, and $9$ qubit problems. Consequentially, the total noise in the circuit grows with more number of qubits and more number of repetitions of amplitude amplification.

\begin{figure}[h!t]
\begin{center}
\includegraphics[width=0.75\linewidth]{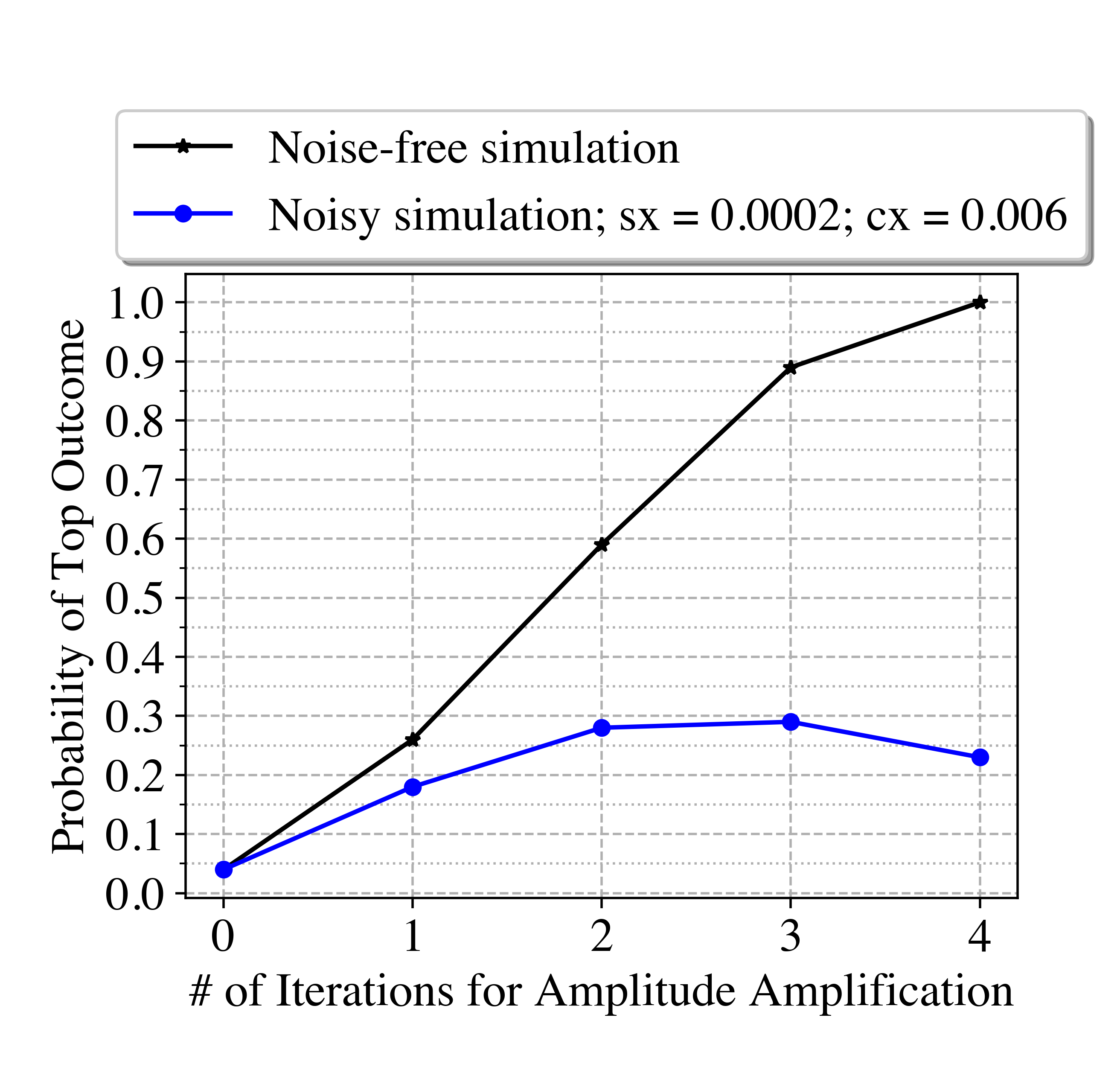}
\end{center}
\caption{Gain from amplitude amplification vs loss from accumulated gate error.}
\label{fig:motivation}
\end{figure}

By amplifying the amplitude of the state vector corresponding to the marked item, the probability of the winning outcome(s) is also amplified almost $3\times$ per iteration in an ideal noise-free environment. However, as the number of quantum gates grows linearly and so thus the total noise in the system, the advantage of amplitude amplification gets diminished. There comes an inflection point when the loss in accuracy from total noise in the system overshadows the gain from repeating the subroutine. Figure~\ref{fig:motivation} demonstrates this reality succinctly. In this figure, we plot probability of the winning outcome against the number of iterations for amplitude amplification for two environments - (i) a completely noise-free simulation, and (ii) an environment mimicking the noise model of IBMQ's \verb|Santiago| quantum resource. We can see for a fact that in noise-free simulation, the outcome probability grows linearly form the initial superposition state marked by the iteration $0$ to the maximum value of $1$ at iteration $4$, the theoretical optimal value. Although, the resulting probability of winning outcome in noisy simulation is always smaller than the same in noise-free simulation, it still is amplified sub-linearly with each repetition from iteration $0$ to $3$ and then drops with another iteration resulting in a convex curve. This demonstrates that, in a noisy environment, it can be beneficial to stop amplification at a point earlier than the one prescribed by the amplitude amplification algorithm.  This point can vary from machine to machine, so this necessitates a framework to predict the optimal number of iterations by considering the noise parameters of the underlying quantum hardware.
\section{Optimal Amplitude Amplification}
\label{sec:scheme}

In this section, we describe the implementation details of a library function \verb|amplify_optimal| that generates the optimal quantum circuit for solving the problem at hand by finding the inflection point.  It works in two parts: a front-end and a back-end.

The front-end generates an amplification circuit in the same way as the pre-existing IBM Qiskit library \verb|Grover| class member function \verb|amplify|, by repeating the the oracle and diffuser circuits illustrated in Figure~\ref{fig:CircuitAmplitudeAmplification}.  The difference is that, while doing so, it instruments the circuit with novel transpiler intrinsics that convey semantics of the amplification code to the back-end for optimization.

The back-end takes the intermediate circuit generated by the front-end and generates a circuit that is optimized for the target back-end quantum machine that is intended to execute this circuit.  When performing optimization, it takes into consideration the information conveyed through intrinsics about the amplification algorithm as well as noise properties of the target machine in deciding where the inflection point is.  It is important that the optimization happens after the intermediate circuit has been mapped to the physical circuit runnable on the machine, taking into consideration available basis gates and geometry, so that the optimizer can make accurate predictions about the noise generated by the circuit.  As such, the back-end is implemented as a transpiler extension.

\subsection{Front-end: Build Circuit Instrumented with Intrinsics}

In order for the back-end transpiler to make an informed decision on whether to proceed with an iteration in the face of noise, first, it must know the amount of amplification achieved per iteration, and second, where the iteration boundaries lie in code.  To this end, we propose four novel intrinsics for the OpenQASM (Open Quantum Assembly Language): $amplification\_begin(\theta)$, $amplification\_end()$, $iteration\_begin()$, and $iteration\_end()$.  These intrinsics are only meant for the consumption of the transpiler and do not translate to quantum gates themselves.  At below is how we would use these intrinsics in actual amplitude amplification code:

\begin{lstlisting}
amplification_begin($\theta$)

iteration_begin()
/* Add oracle circuit for iteration 1. */
/* Add diffuser circuit for iteration 1. */
iteration_end()
iteration_begin()
/* Add oracle circuit for iteration 2. */
/* Add diffuser circuit for iteration 2. */
iteration_end()
...
iteration_begin()
/* Add oracle circuit for iteration $t_{opt}$. */
/* Add diffuser circuit for iteration $t_{opt}$. */
iteration_end()

amplification_end
\end{lstlisting}

The $\theta$ is the rotational angle used in each amplification step.  Iteration number $t_{opt}$ is the iteration when $\lvert w \rangle$, the winning state, is reached in the absence of noise.  The $iteration\_begin()$ and $iteration\_end()$ and delineate iteration boundaries. Our \verb|amplify_optimal| generates the above code after computing $\theta$ and $t_{opt}$ as shown in Method~\ref{alg:amplify_optimal}.

\begin{algomethod}
\caption{Build amplification circuit instrumented with intrinsics}
\label{alg:amplify_optimal}
\SetAlgoLined
\DontPrintSemicolon
\SetKwProg{Def}{def}{:}{}
\Def{amplify\_optimal(m: int, n: int)}{
    \tcc{$m$: number of solutions \\
    $n$: number of qubits}
    \;
    $\text{amplitude} \gets \sqrt{\frac{m}{2^n}}$\;
    $\theta \gets \arcsin{(\text{amplitude})}$\;
    \;
    $t_{opt} \gets \frac{\arccos{(\text{amplitude})}}{2*\arcsin(\text{amplitude})}$\;
    \;
    \tcp{Use $\theta$ and $t_{opt}$ to build circuit}
    $C \gets \text{circuit instrumented with intrinsics}$\;
    \KwRet $C$\;
}
\end{algomethod}

As described in Section~\ref{subsec:amp_amplification}, the amplitude amplification process starts with an initial superposition state $\lvert s \rangle = \sin(\theta)\lvert w \rangle + \cos(\theta)\lvert s' \rangle$, where $\theta = \arcsin(\langle s \lvert w \rangle)$. For a single solution or one marked item in $N=2^n$ dimensional solution space, $\theta = \arcsin(\langle s \lvert w \rangle) = \arcsin(\frac{1}{\sqrt{N}})$. It can be further generalized to show that for a problem with $M$ solutions, the initial amplitude can be expressed as $\theta = \arcsin{(\sqrt{\frac{M}{N}})}$.

As described in Section~\ref{subsec:amp_amplification}, amplitude amplification is implemented as an oracle circuit followed by a diffuser circuit, which can be expressed as the transformation $U_{s}U_{f}$. To achieve a higher probability, this process is repeated several times as shown in Figure~\ref{fig:CircuitAmplitudeAmplification}. The resulting state of the marked item after $t$ iterations can be expressed as $\psi_t = (U_{s}U_{f})^{t}\lvert s \rangle = \sin(\theta_{t})\lvert w \rangle + \cos(\theta_{t})\lvert s' \rangle$, where $\theta_{t} = (2t + 1)\theta$. The process of iterating over oracle and diffuser circuit stops when $\psi_t$ overlaps with the winning state, $\lvert w \rangle$. Geometrically, this means $\theta_{t} = (2t + 1)\theta = \frac{\pi}{2}$. By using trigonometric identity, $\frac{\pi}{2} = \arcsin{(\theta)} + \arccos{(\theta)}$, the optimal number of iterations to attain highest outcome probability can be expressed as $t_{opt} = \frac{\arccos{(\sqrt{\frac{M}{N}})}}{2*\arcsin{(\sqrt{\frac{M}{N}})}}$. This mathematical formula is captured in Method~\ref{alg:amplify_optimal}.

\subsection{Back-end: Model Depolarizing Noise Channel}

Given amplification information through the intrinsics, now the transpiler calculates a prediction for the degradation in the winning outcome due to noise.  In order to do that, we need to return to Equation~\ref{eq:depolarizing_channel} for the depolarizing noise channel. The mixed state $\rho$ in the equation is expressed as a density matrix, which is an ensemble of multiple pure states, each with an associated probability of occurrence.  Since the probabilities of occurrence for the basis states always add up to 1, the trace of the density matrix, $Tr[\rho] = 1$.  Hence, the equation can be rewritten as:

\begin{equation}
\label{eq:depolarizing_channel_simplified}
    \begin{aligned}
        E_{\lambda}(\rho)&=(1-\lambda)\rho + {\lambda}{\frac{I}{2^n}} \\
        \text{with } & 0 \leq \lambda \leq \frac{4^n}{4^n - 1} \\
        \text{where, } &\\
        \lambda & \text{ is the depolarizing error parameter} \\ 
        n & \text{ is the number of qubits.} \\
    \end{aligned}
\end{equation}

\begin{figure}[htp]
    \centering
    \subcaptionbox{With probability $1 - \lambda$, no change is made.\label{fig:single_qubit_depolarization_sx}}[0.7\linewidth]
        {\includegraphics[width=0.8\linewidth]{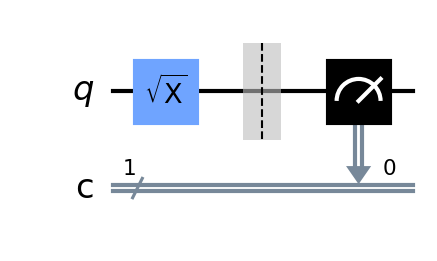}\vspace{-0.2in}}
    \subcaptionbox{With probability $\lambda / 2$, state is basis state $\lvert 0 \rangle$.\label{fig:single_qubit_depolarization_0}}[0.7\linewidth]
        {\includegraphics[width=0.8\linewidth]{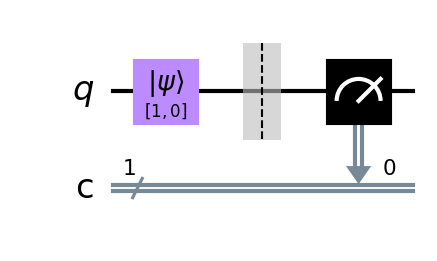}\vspace{-0.2in}}
    \subcaptionbox{With probability $\lambda / 2$, state is basis state $\lvert 1 \rangle$.\label{fig:single_qubit_depolarization_1}}[0.7\linewidth]
        {\includegraphics[width=0.8\linewidth]{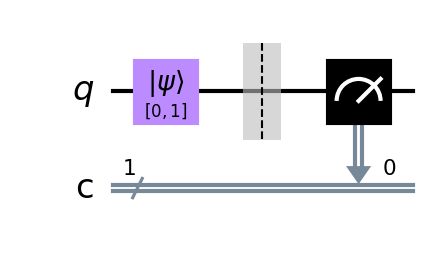}\vspace{-0.2in}}
    \caption{Three states in the ensemble when depolarizing error is applied to a single qubit gate sx.}
    \label{fig:single_qubit_depolarization}
\end{figure}

To wrap our heads around this equation, let us discuss an example where $n =
1$ and a depolarizing noise channel is applied to a single qubit sx gate, where the circuit looks like Figure~\ref{fig:single_qubit_depolarization_sx}.  Since $I = \lvert 0 \rangle\langle 0 \lvert + \lvert 1 \rangle\langle 1 \lvert$, $\rho$ can be expressed as an ensemble of three states: the original state (Figure~\ref{fig:single_qubit_depolarization_sx}), the basis state $\lvert 0 \rangle$ (Figure~\ref{fig:single_qubit_depolarization_0}), and the basis state $\lvert 1 \rangle$ (Figure~\ref{fig:single_qubit_depolarization_1}), with probabilities $1 - \lambda$, $\lambda / 2$, and $\lambda / 2$ respectively.

When $\lambda = 0$, that is when there is no error, the sx gate is unaffected and always functions like the original circuit ~\ref{fig:single_qubit_depolarization_sx}.  When $\lambda = 1$, $\rho$ becomes a maximally mixed state $\frac{1}{2} (\lvert 0 \rangle\langle 0 \lvert + \lvert 1 \rangle\langle 1 \lvert)$, where there is equal probability of the circuit functioning like ~\ref{fig:single_qubit_depolarization_0} or ~\ref{fig:single_qubit_depolarization_1}.

\begin{figure}[htp]
    \centering
    \subcaptionbox{With probability $1 - \lambda$, no change is made.\label{fig:two_qubit_depolarization_cx}}
        {\includegraphics[width=0.7\linewidth]{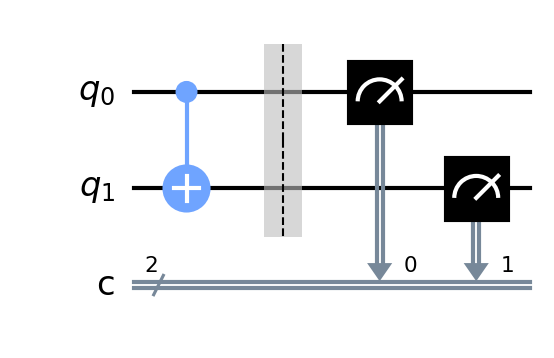}}
    \subcaptionbox{With probability $\lambda / 4$, state is basis state $\lvert 00 \rangle$.\label{fig:two_qubit_depolarization_00}}
        {\includegraphics[width=0.7\linewidth]{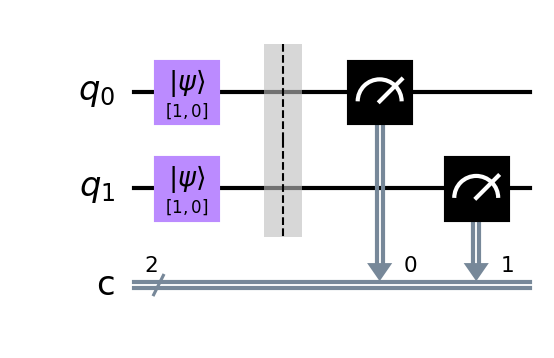}}
    \subcaptionbox{With probability $\lambda / 4$, state is basis state $\lvert 01 \rangle$.\label{fig:two_qubit_depolarization_01}}
        {\includegraphics[width=0.7\linewidth]{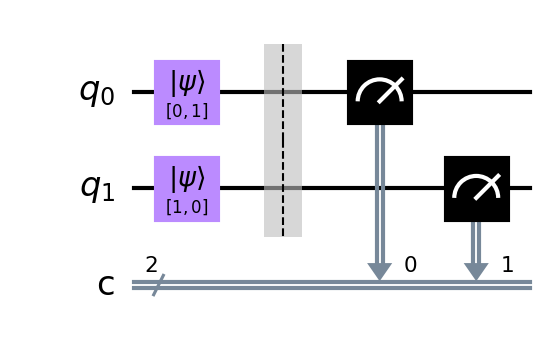}}
    \subcaptionbox{With probability $\lambda / 4$, state is basis state $\lvert 10 \rangle$.\label{fig:two_qubit_depolarization_10}}
        {\includegraphics[width=0.7\linewidth]{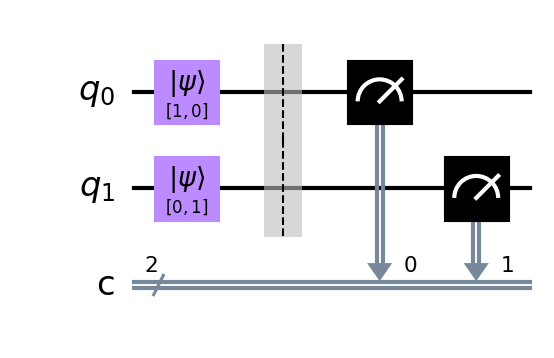}}
    \subcaptionbox{With probability $\lambda / 4$, state is basis state $\lvert 11 \rangle$.\label{fig:two_qubit_depolarization_11}}
        {\includegraphics[width=0.7\linewidth]{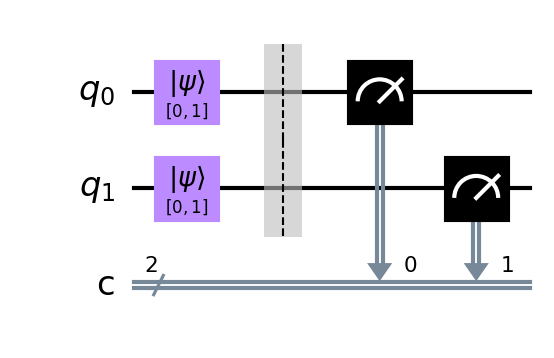}}
    \caption{Five states in the ensemble when depolarizing error is applied to a two qubit gate cx.}
    \label{fig:two_qubit_depolarization}
\end{figure}

When $n = 2$ and a depolarizing noise channel is applied to a two qubit gate such as cx gate, the ensemble of states in $\rho$ looks like Figure~\ref{fig:two_qubit_depolarization}.  In this case, $I = \lvert 00 \rangle\langle 00 \lvert + \lvert 01 \rangle\langle 01 \lvert + \lvert 10 \rangle\langle 10 \lvert + \lvert 11 \rangle\langle 11 \lvert$.  So, we will obtain an ensemble of five states: the original state (Figure~\ref{fig:two_qubit_depolarization_cx}), the basis state $\lvert 00 \rangle$ (Figure~\ref{fig:two_qubit_depolarization_00}), the basis state $\lvert 01 \rangle$ (Figure~\ref{fig:two_qubit_depolarization_01}), the basis state $\lvert 10 \rangle$ (Figure~\ref{fig:two_qubit_depolarization_10}), and the basis state $\lvert 11 \rangle$ (Figure~\ref{fig:two_qubit_depolarization_11}).

We verified our understanding of the depolarizing error channel by
experimenting with many microcircuits on IBM Qiskit.  We verified that the
outcome of a circuit when applied depolarizing noise is actually equivalent to
the sum of the outcomes of the ensemble circuits weighted by probability, as
described in Figure~\ref{fig:single_qubit_depolarization} and
Figure~\ref{fig:two_qubit_depolarization}.

\subsection{Back-end: Calculate Error Probability from Parameter}

Given our understanding of the depolarizing error channel now, we would like to calculate error probability $P$ from depolarizing error parameter $\lambda$.  It is easy to mistakenly think that $P = \lambda$ from the construction of Equation~\ref{eq:depolarizing_channel_simplified}.  However, things are not that simple.

Let's again start by reasoning about depolarizing error for $n = 1$.  $P$ denotes the probability that a random rotation is performed on the qubit due to depolarizing noise.  The rotation can happen with equal probability on the X, Y, and Z axes in the Bloch sphere, giving us the equation:

\begin{equation}
\label{eq:depolarizing_channel_pauli}
    \begin{aligned}
        E_{P}(\rho)&=(1 - P)\rho + \frac{P}{3}(X \rho X + Y \rho Y + Z \rho Z)
    \end{aligned}
\end{equation}

Now, how is Equation~\ref{eq:depolarizing_channel_pauli} related to Equation~\ref{eq:depolarizing_channel_simplified} which is parameterized by $\lambda$ instead of $P$?  We know that by Pauli twirling we can achieve the maximally mixed state $\frac{I}{2}$:

\begin{equation}
\label{eq:pauli_twirling}
    \begin{aligned}
        \frac{I}{2}&=\frac{1}{4}(I \rho I + X \rho X + Y \rho Y + Z \rho Z)
    \end{aligned}
\end{equation}

If we apply Equation~\ref{eq:pauli_twirling} to Equation~\ref{eq:depolarizing_channel_pauli}:

\begin{equation}
\label{eq:depolarizing_channel_derivation}
    \begin{aligned}
        E_{P}(\rho) &= (1 - P)\rho + \frac{P}{3}(X \rho X + Y \rho Y + Z \rho Z) \\
        &= (1 - \frac{4}{3}P)\rho + \frac{P}{3}(I \rho I + X \rho X + Y \rho Y + Z \rho \ Z) \\
        &= (1 - \frac{4}{3}P)\rho + \frac{4}{3}P * {\frac{I}{2}}
    \end{aligned}
\end{equation}

Comparing that to $E_{\lambda}(\rho)=(1-\lambda)\rho + {\lambda}{\frac{I}{2^n}}$ in Equation~\ref{eq:depolarizing_channel_simplified}, it is easy to derive the equality:

\begin{equation}
\label{eq:P_expression_for_one_qubit}
    \begin{aligned}
        E_{P}(\rho) &= E_{\lambda}(\rho) \Rightarrow \lambda = \frac{4}{3}{P} \Rightarrow P = \frac{3}{4}{\lambda}
    \end{aligned}
\end{equation}

Now Equation~\ref{eq:depolarizing_channel_pauli} can be generalized to an arbitrary $n$:

\begin{equation}
\label{eq:depolarizing_channel_general}
    \begin{aligned}
        E_{P}(\rho) &= (1 - P)\rho + \frac{P}{4^n-1}(\sum_{j}{P_{j} \rho P_{j}} \ for \ all \ P_{j} != I)
    \end{aligned}
\end{equation}

Also, the Pauli twirl can be generalized to $n$ qubits as well:

\begin{equation}
\label{eq:general_pauli_twirling}
    \begin{aligned}
        \frac{I}{2^n}&=\frac{1}{4^n}(\sum_{j}{P_{j} \rho P_{j}})
    \end{aligned}
\end{equation}

Applying Equation~\ref{eq:general_pauli_twirling} to Equation~\ref{eq:depolarizing_channel_general}, the following is derived:

\begin{equation}
\label{eq:depolarizing_channel_general_derivation}
    \begin{aligned}
        E_{P}(\rho) &= (1 - P)\rho + \frac{P}{4^n-1}(\sum_{j}{P_{j} \rho P_{j}} \ for \ all \ P_{j} != I) \\
        &= (1 - \frac{4^n}{4^n-1}P)\rho + \frac{P}{4^n-1}(\sum_{j}{P_{j} \rho P_{j}}) \\
        &= (1 - \frac{4^n}{4^n-1}P)\rho + \frac{4^n}{4^n-1}P * {\frac{I}{2^n}} \\
        \text{where, } &\\
        P_{j} & \text{ is all permutations of Pauli channels for all qubits}
    \end{aligned}
\end{equation}

Finally, comparing Equation~\ref{eq:depolarizing_channel_general_derivation} against Equation~\ref{eq:depolarizing_channel_simplified}, we derive generally for $n$ the value of $P$:

\begin{equation}
\label{eq:P_expression_for_n_qubits}
    \begin{aligned}
        E_{P}(\rho) &= E_{\lambda}(\rho) \Rightarrow \lambda = \frac{4^n}{4^n-1}P \Rightarrow P = \frac{4^n-1}{4^n}{\lambda} 
    \end{aligned}
\end{equation}

\subsection{Back-end: Calculate Noise per Iteration}

A quantum machine back-end advertises the noise parameters ($\lambda$s) for each physical gate in the machine in a long table.  From these $\lambda$s, we can apply Equation~\ref{eq:P_expression_for_n_qubits} to convert them to individual noise rates.

Our transpiler extension uses these noise rates to calculate the noise for the given iteration circuit as shown in Method~\ref{alg:noise_free_probability}. It returns the amount of degradation in probability suffered by the winning outcome due to noise for the given iteration.  The function $calculate\_noise$ takes a circuit $C$ as an argument which is the circuit between the $iteration\_begin$ and $iteration\_end$ intrinsics. The circuit would contain the oracle and diffuser transformations for that iteration.

The function loops over all gates $g$ in the circuit and retrieves the corresponding $\lambda$ from the aforementioned long table and uses Equation~\ref{eq:P_expression_for_n_qubits} to convert it to a noise rate $p$, substituting $n$ for 1 or 2 depending on whether $g$ is a one-qubit-gate or a two-qubit-gate. Next, we use Bayes' theorem to add $p$ to the total $noise\_rate$ for the given incremental circuit $C$.

\begin{algomethod}
\caption{Calculate loss of accuracy due to noise.}
\label{alg:noise_free_probability}
\SetAlgoLined
\DontPrintSemicolon
\SetKwProg{Def}{def}{:}{}
\Def{calculate\_noise(C: circuit)}{
    $noise\_rate \gets 0$\;
    \;
    \tcp{Loop over all gates to account for their noise}
    \For{each gate $g$ in $C$}{
    
        $\lambda \gets \text{noise parameter for gate g}$\;
        \;
        \tcp{Calculate noise probabilities from noise parameters}
        \If{$g$ is a one-qubit-gate}{
            $p \gets \lambda*\frac{4^1 - 1}{4^1}$\;
        }
        \If{$g$ is a two-qubit-gate}{
            $p \gets \lambda*\frac{4^2 - 1}{4^2}$\;
        }
        
        \;
        $noise\_rate \gets 1 - (1 - noise\_rate) * (1 - p)$\;
    }
    \;
    
    \KwRet $noise\_rate$\;
}
\end{algomethod}

\subsection{Back-end: Predict Inflection Point and Optimize}

Now, we are finally able to predict the inflection point where further iterations do not increase accuracy. In other words, we are trying to find the optimal number of iterations where the probability of the marked item is the highest. Function $optimize\_circuit$ in Method~\ref{alg:estimated_optimal} takes as arguments: circuit $C$, the entire amplification circuit between the $amplification\_begin(\theta)$ and $amplification\_end()$ intrinsics, and $\theta$ which is the rotational angle passed to $amplification\_begin(\theta)$.  $optimize\_circuit$ loops over all iterations of circuit $C$ until it finds the inflection point.  We use variable $t$ to keep track of our current iteration.

\begin{algomethod}
\caption{Predict inflection point when noise for a certain iteration starts to trump amplification gains.}
\label{alg:estimated_optimal}
\SetAlgoLined
\DontPrintSemicolon
\SetKwProg{Def}{def}{:}{}
\Def{optimize\_circuit(C: circuit, $\theta$: float)}{
    \tcc{$C$: amplification circuit \\
    $\theta$: rotational angle}
    \;
    \tcp{$noise$ is cumulative noise}
    $noise \gets 0$\;
    \;
    \tcp{$t$ is current iteration number}
    $t \gets 1$\;
    \;
    \tcp{Loop on iterations in $C$}
    \While{\text{no more iterations in $C$}}{
        $amplification \gets \sin^2{((2t+1)*\theta)}$\;
        \;
        \tcp{$C_t$ is circuit for iteration $t$}
        $C_t \gets extract\_iteration\_circuit(C, t)$\;
        \;
        \tcp{$noise_t$ is noise for iteration $t$}
        $noise_t \gets calculate\_noise(C_t)$\;
        \;
        $noise \gets 1 - (1 - noise) * (1 - noise_t)$\;
        \;
        \If{$amplification \leq noise$}{
            $C \gets \text{$C$ with $C_r$ removed where $r \geq t$}$
            \KwRet  $C$\;
        }
        \;
        $t \gets t+1$\;
    }
    \;
    \KwRet  $C$\;
}
\end{algomethod}

At each iteration, first we calculate $amplification$, which is the probability of the winning outcome after $t$ iterations, in a noiseless environment.  Recall from Section~\ref{subsec:amp_amplification} that the probability of the state corresponding to the marked item being collapsed on the winning state, $\lvert w \rangle$, during the measurement is the projection of the state vector on the $\lvert w \rangle$ axis at any given time. Thus,
$amplification = \sin^2{\theta_t} = \sin^2{(2t+1)\theta} = \sin^2{((2t+1)*\arcsin{(\sqrt{\frac{M}{2^n}})})}$.

Next the circuit for iteration $t$ is extracted from $C$ and assigned to $C_t$.  We use $C_t$ to calculate $noise_t$ incurred from that iteration using Method~\ref{alg:noise_free_probability}, which we accumulate into $noise$, again using Bayes' theorem.

 When $noise$ trumps $amplification$, we have found our inflection point.  At that point, we remove all remaining iterations from circuit $C$ and then return the optimized circuit.
 
Note that there is a potential for inaccuracy in Method~\ref{alg:estimated_optimal} however.  We conservatively assumed that none of the noisy results included in $noise$ would end up in the winning state.  Some noisy results may in fact end up in the winning state.  But we will show in our results that this inaccuracy is minimal.
\section{Experimental Evaluation}
\label{sec:results}

This section describes the implementation framework of the proposed scheme followed by a detailed analysis of the experimental results.

\subsection{Framework}
\label{subsec:framework}

\begin{figure*}[h!t]
    \centering
    \subcaptionbox{Single-qubit gate noise\label{fig:9q_single_result}}
        {\includegraphics[width=0.32\linewidth]{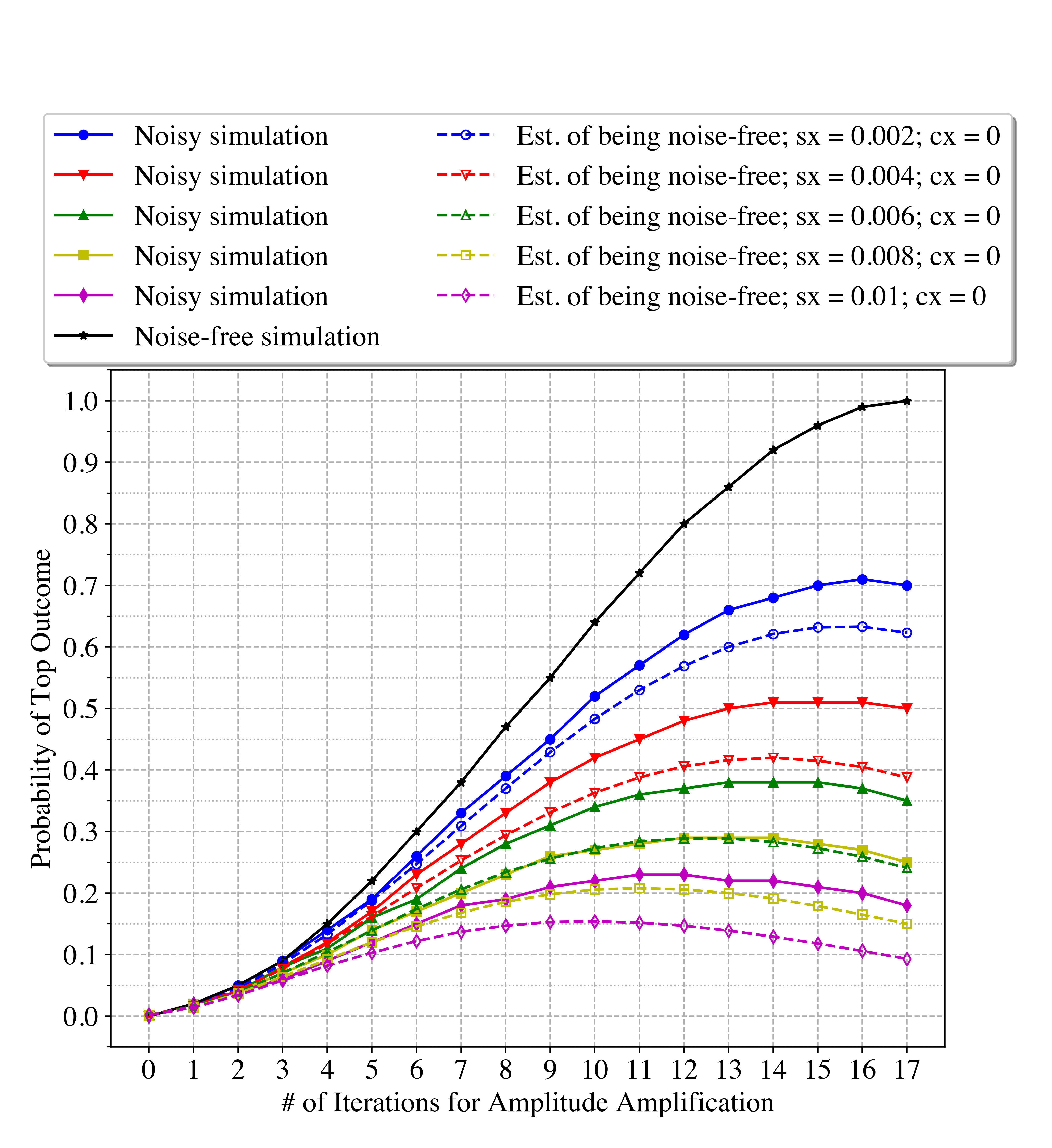}}
    \subcaptionbox{Two-qubit gate noise\label{fig:9q_two_result}}
        {\includegraphics[width=0.32\linewidth]{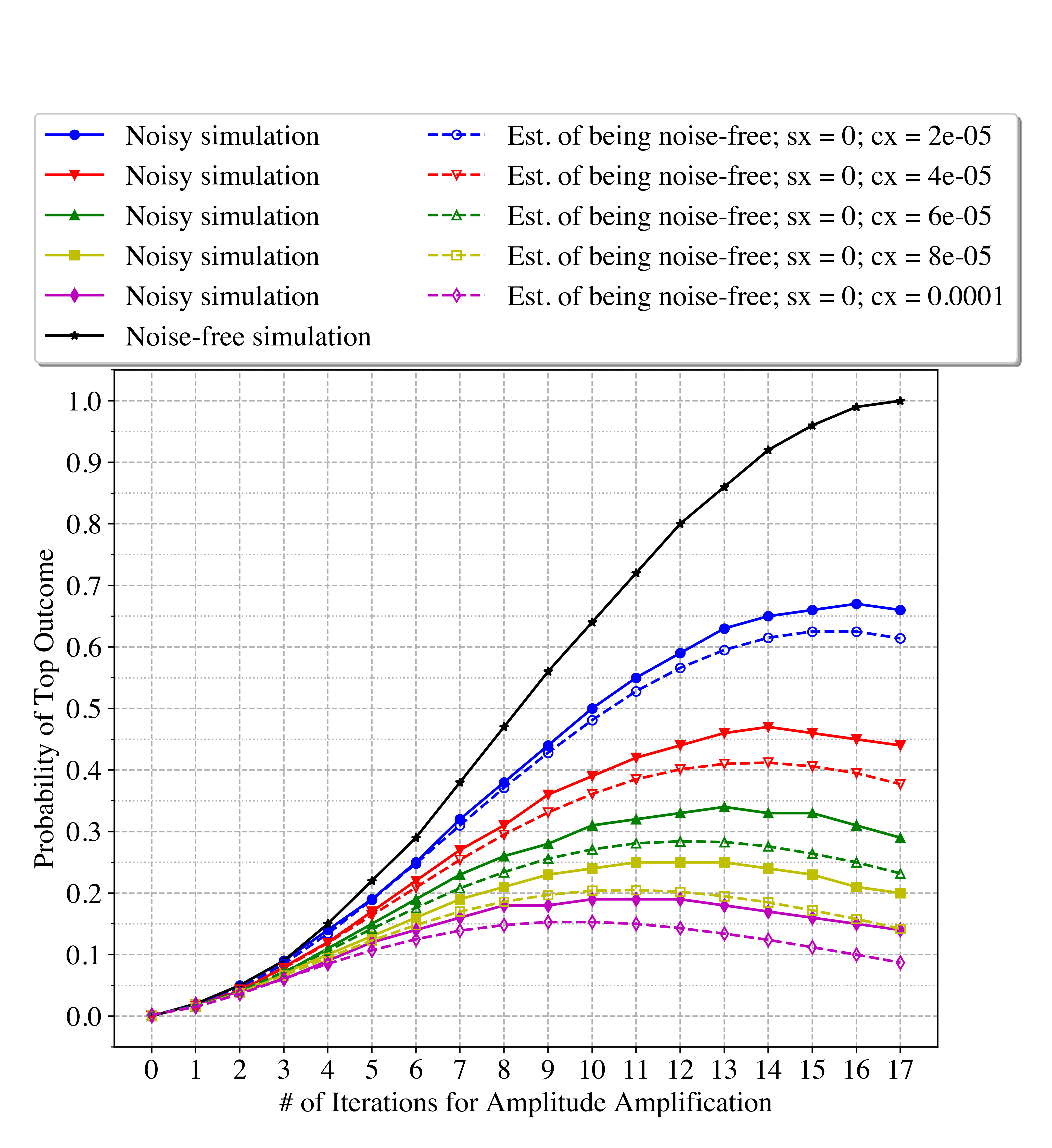}}
    \subcaptionbox{Both single and two-qubit gate noise\label{fig:9q_both_result_gt}}
        {\includegraphics[width=0.32\linewidth]{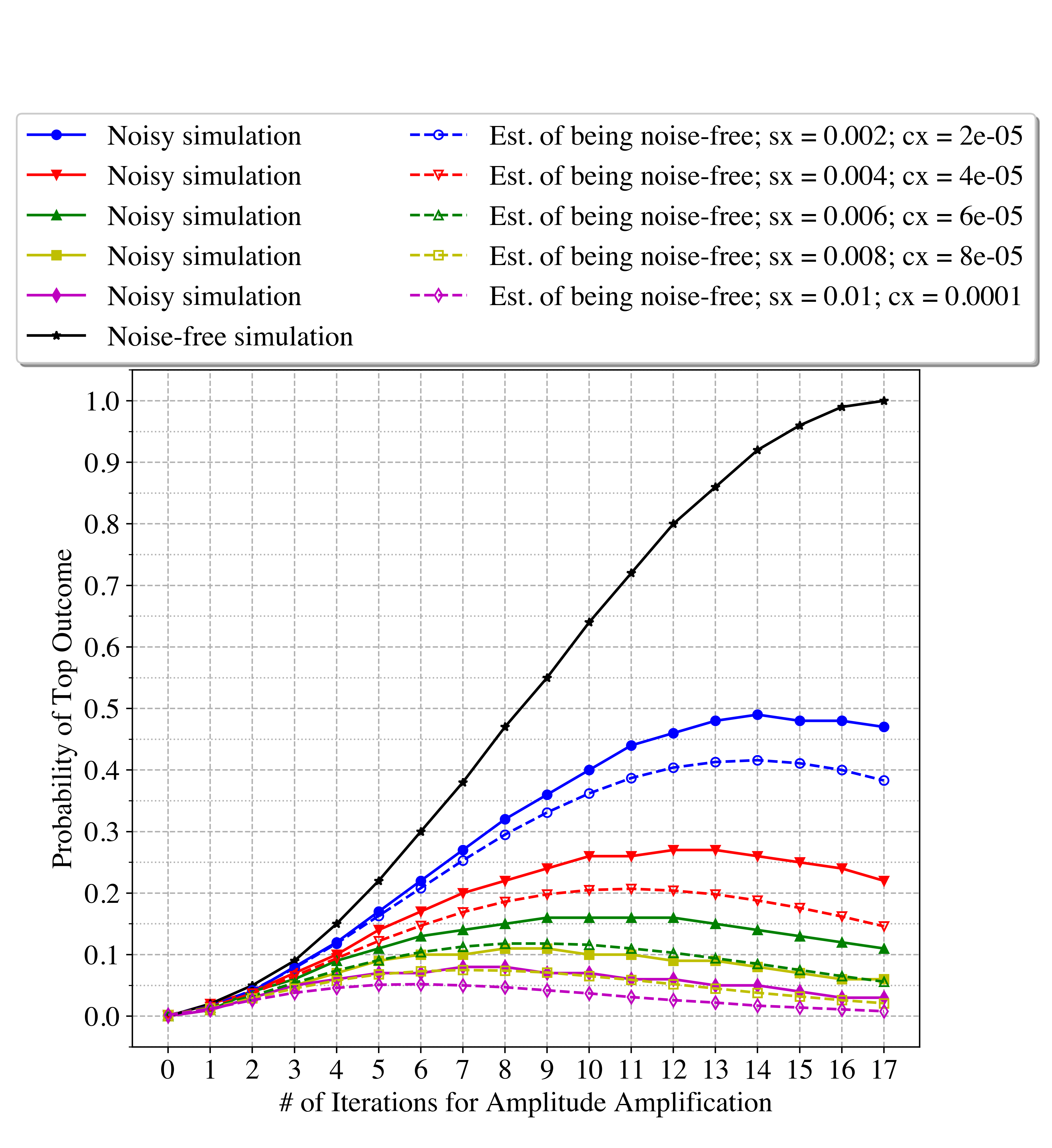}}
    \caption{Effect of varied types of gate error on quantum circuits.}
    \label{fig:gate_type_variation_result}
\end{figure*}

\begin{figure*}[h!t]
    \centering
    \subcaptionbox{5-qubit circuit\label{fig:5q_both_result}}
        {\includegraphics[width=0.32\linewidth]{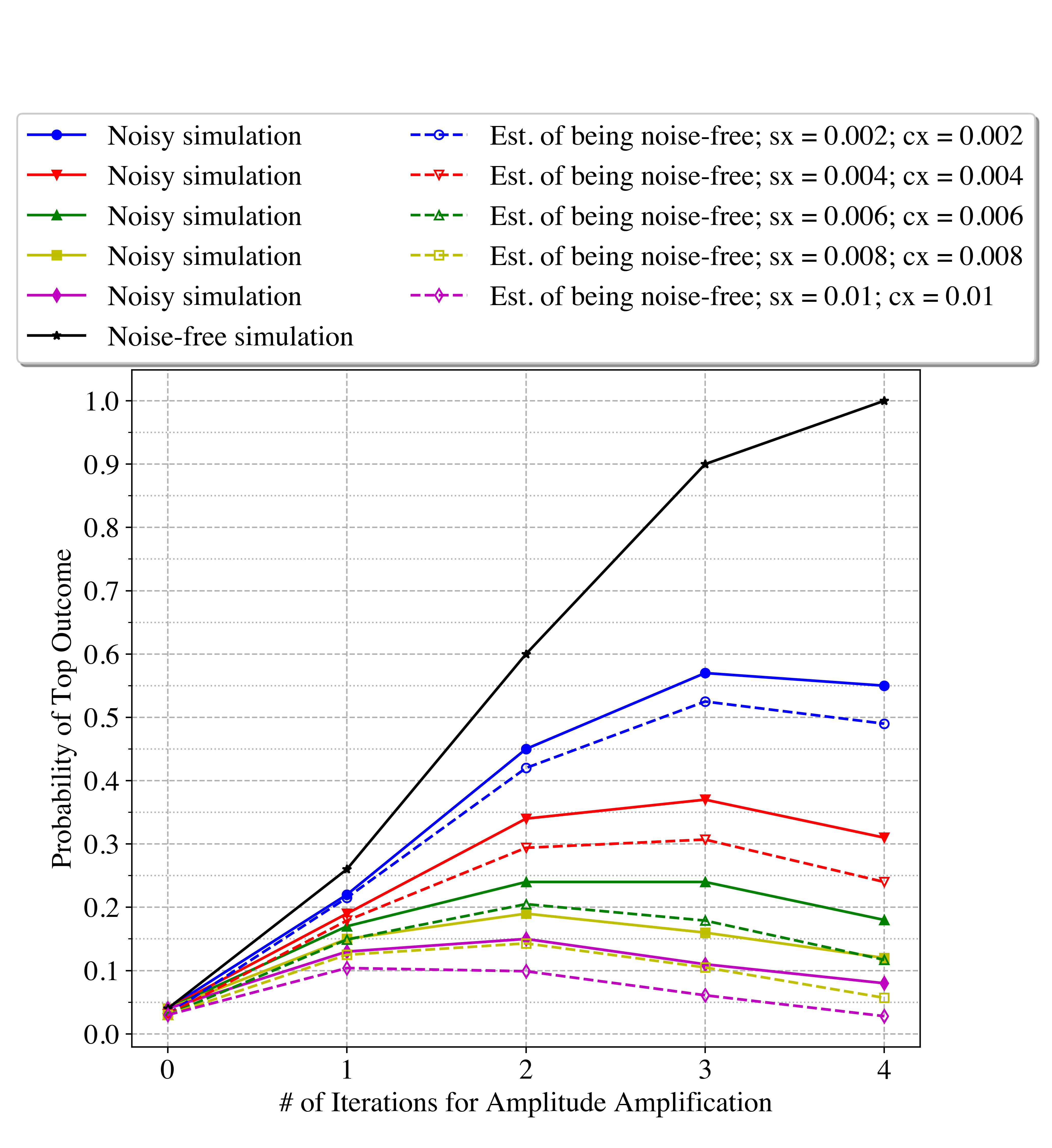}}
    \subcaptionbox{7-qubit circuit\label{fig:7q_both_result}}
        {\includegraphics[width=0.32\linewidth]{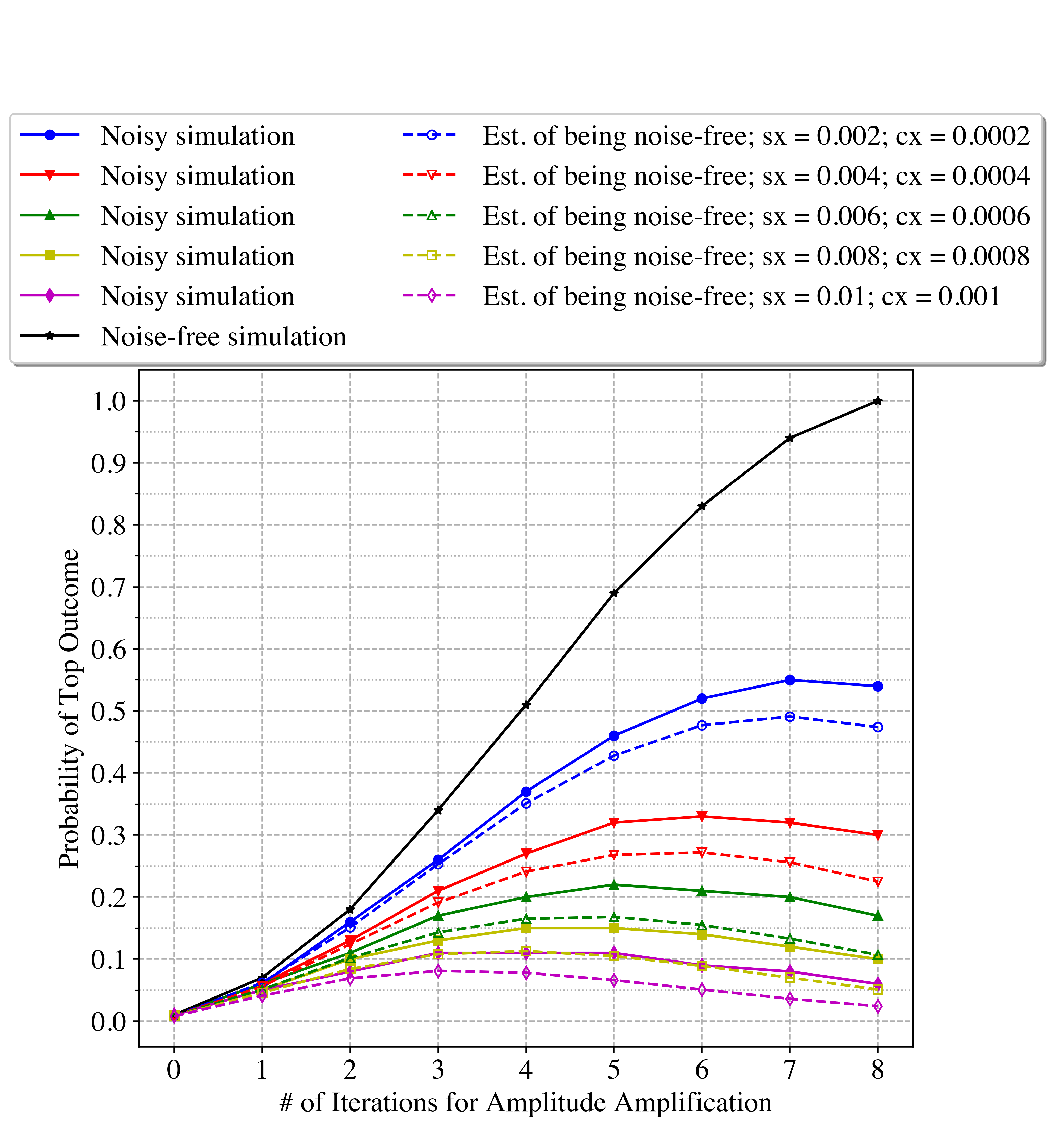}}
    \subcaptionbox{9-qubit circuit\label{fig:9q_both_result}}
        {\includegraphics[width=0.32\linewidth]{Images/9q_both_result.png}}
    \caption{Effect of single and two-qubit gate noise on circuits with varied number of qubits.}
    \label{fig:num_qubit_variation_result}
\end{figure*}

The methods presented in Section~\ref{sec:scheme} are implemented as part of \verb|AmplitudeAmplifier| and \verb|Grover| class of \verb|IBM Qiskit v0.20.2| library~\cite{IBMQiskit}. To evaluate the efficacy of the method predicting the optimal iteration, we used Grover's amplitude amplification algorithm to solve a Boolean satisfiability (SAT) problem. We will publish the concrete implementation of the methods described in Section~\ref{sec:scheme} along with the SAT solver quantum workloads upon acceptance. The test setup allows the flexibility to vary the number of qubits, $n$ and the number of solutions, $M$ to compose the SAT solver.  Users can also pass as input argument the marked item(s) for the SAT solver.  We specifically experimented with 1-SAT and 3-SAT problem by setting $M$ to 1 and 3 respectively. We will focus on 1-SAT mostly and report three variations of 1-SAT with different numbers of qubits, for $n=5$, $n=7$, and $n=9$.  This is done to test our framework on different circuit sizes as shown in Table~\ref{tab:gate_count}.  We will show results for 3-SAT along with an Un-Ordered Search (UOS) benchmark later in this section.

IBM Qiskit provides a \verb|PhaseOracle| class simplifying the creation of an oracle circuit from a Boolean expression and an \verb|AmplificationProblem| class to compose a SAT-solver problem from the created oracle. The \verb|Grover| class, which extends the general \verb|AmplitudeAmplifier|, can be instantiated and then its \verb|amplify| method used to amplify the amplitude of the item marked by the oracle. The \verb|Grover| object can be instantiated to specify the number of iterations that the Grover operator should be applied. Users can also specify a \verb|QuantumInstance|, while instantiating the \verb|Grover| object, which is used to execute the transpiled circuit for a specified iteration number.  We report results of execution on a noise-free \verb|aer_simulator| backend and a \verb|qasm_simulator| backend to model a noisy quantum resource. Both of these backends are built while limiting the basis gates to an identical set composed of \verb|sx|, \verb|rz|, and \verb|cx| gates, such that both will produce identical transpiled circuits given the same quantum program.

To model a noisy quantum resource, we used a quantum depolarizing channel. When creating the noise model for the \verb|qasm_simulator| backend, the noise parameters are set for \verb|sx|, \verb|rz|, and \verb|cx| basis gates as the transpiled circuit is composed of these gates alone.  To reduce variability in the results, we specified a sufficiently large value for the number of shots while creating the backend. We repeatedly ran each experiment several times to verify that the variability is indeed negligible and we have a statistically meaningful result.


\subsection{Analysis of Experimental Results}
\label{subsec:analysis_result}

\begin{table*}[h!t]
\caption{Optimal number of iterations observed in noisy simulation and predicted by our method along with their observed outcome probability}
\label{tab:result}
\begin{adjustbox}{width=\textwidth,keepaspectratio}
\begin{tabular}{|c|ll|l|l|l|ll|}
\hline
\multirow{3}{*}{\textbf{\begin{tabular}[c]{@{}c@{}}\# of \\ \\ qubits\end{tabular}}} & \multicolumn{2}{c|}{\multirow{2}{*}{\textbf{Gate Error Rate}}}      & \multicolumn{1}{c|}{\multirow{3}{*}{\textbf{\begin{tabular}[c]{@{}c@{}}Theoretical \\ optimal \#\\ of iterations\end{tabular}}}} & \multicolumn{1}{c|}{\multirow{3}{*}{\textbf{\begin{tabular}[c]{@{}c@{}}Observed optimal \# \\ of iterations in \\ simulated env.\end{tabular}}}} & \multicolumn{1}{c|}{\multirow{3}{*}{\textbf{\begin{tabular}[c]{@{}c@{}}Predicted\\ optimal \# \\ of iterations\end{tabular}}}} & \multicolumn{2}{c|}{\multirow{2}{*}{\textbf{\begin{tabular}[c]{@{}c@{}}Success probability\\ in simulated environment\end{tabular}}}}                                                                   \\
                                                                                     & \multicolumn{2}{c|}{}                                               & \multicolumn{1}{c|}{}                                                                                                            & \multicolumn{1}{c|}{}                                                                                                                              & \multicolumn{1}{c|}{}                                                                                                              & \multicolumn{2}{c|}{}                                                                                                                                                                                             \\ \cline{2-3} \cline{7-8} 
                                                                                     & \multicolumn{1}{c|}{\textbf{sx}} & \multicolumn{1}{c|}{\textbf{cx}} & \multicolumn{1}{c|}{}                                                                                                            & \multicolumn{1}{c|}{}                                                                                                                              & \multicolumn{1}{c|}{}                                                                                                              & \multicolumn{1}{c|}{\textbf{\begin{tabular}[c]{@{}c@{}}For optimal power\\ in simulated env.\end{tabular}}} & \multicolumn{1}{c|}{\textbf{\begin{tabular}[c]{@{}c@{}}For estimated\\ optimal power\end{tabular}}} \\ \hline
\multirow{5}{*}{\textbf{5}}                                                          & \multicolumn{1}{l|}{0.002}       & 0.002                            & \multirow{5}{*}{4}                                                                                                               & 3                                                                                                                                                  & 3                                                                                                                                  & \multicolumn{1}{l|}{0.57}                                                                                   & 0.57                                                                                                \\ \cline{2-3} \cline{5-8} 
                                                                                     & \multicolumn{1}{l|}{0.004}       & 0.004                            &                                                                                                                                  & 3                                                                                                                                                  & 3                                                                                                                                  & \multicolumn{1}{l|}{0.37}                                                                                   & 0.37                                                                                                \\ \cline{2-3} \cline{5-8} 
                                                                                     & \multicolumn{1}{l|}{0.006}       & 0.006                            &                                                                                                                                  & 2                                                                                                                                                  & 2                                                                                                                                  & \multicolumn{1}{l|}{0.24}                                                                                   & 0.24                                                                                                \\ \cline{2-3} \cline{5-8} 
                                                                                     & \multicolumn{1}{l|}{0.008}       & 0.008                            &                                                                                                                                  & 2                                                                                                                                                  & 2                                                                                                                                  & \multicolumn{1}{l|}{0.19}                                                                                   & 0.19                                                                                                \\ \cline{2-3} \cline{5-8} 
                                                                                     & \multicolumn{1}{l|}{0.01}        & 0.01                             &                                                                                                                                  & 2                                                                                                                                                  & 1                                                                                                                                  & \multicolumn{1}{l|}{0.15}                                                                                   & 0.13                                                                                                \\ \hline
\multirow{5}{*}{\textbf{7}}                                                          & \multicolumn{1}{l|}{0.002}       & 0.0002                           & \multirow{5}{*}{8}                                                                                                               & 7                                                                                                                                                  & 7                                                                                                                                  & \multicolumn{1}{l|}{0.55}                                                                                   & 0.55                                                                                                \\ \cline{2-3} \cline{5-8} 
                                                                                     & \multicolumn{1}{l|}{0.004}       & 0.0004                           &                                                                                                                                  & 6                                                                                                                                                  & 5                                                                                                                                  & \multicolumn{1}{l|}{0.33}                                                                                   & 0.32                                                                                                \\ \cline{2-3} \cline{5-8} 
                                                                                     & \multicolumn{1}{l|}{0.006}       & 0.0006                           &                                                                                                                                  & 5                                                                                                                                                  & 4                                                                                                                                  & \multicolumn{1}{l|}{0.22}                                                                                   & 0.20                                                                                                \\ \cline{2-3} \cline{5-8} 
                                                                                     & \multicolumn{1}{l|}{0.008}       & 0.0008                           &                                                                                                                                  & 4                                                                                                                                                  & 3                                                                                                                                  & \multicolumn{1}{l|}{0.15}                                                                                   & 0.13                                                                                                \\ \cline{2-3} \cline{5-8} 
                                                                                     & \multicolumn{1}{l|}{0.01}        & 0.001                            &                                                                                                                                  & 3                                                                                                                                                  & 3                                                                                                                                  & \multicolumn{1}{l|}{0.11}                                                                                   & 0.11                                                                                                \\ \hline
\multirow{15}{*}{\textbf{9}}                                                         & \multicolumn{1}{l|}{0.002}       & 0                                & \multirow{15}{*}{17}                                                                                                             & 16                                                                                                                                                 & 15                                                                                                                                 & \multicolumn{1}{l|}{0.71}                                                                                   & 0.70                                                                                                \\ \cline{2-3} \cline{5-8} 
                                                                                     & \multicolumn{1}{l|}{0.004}       & 0                                &                                                                                                                                  & 14                                                                                                                                                 & 13                                                                                                                                 & \multicolumn{1}{l|}{0.51}                                                                                   & 0.50                                                                                                \\ \cline{2-3} \cline{5-8} 
                                                                                     & \multicolumn{1}{l|}{0.006}       & 0                                &                                                                                                                                  & 13                                                                                                                                                 & 12                                                                                                                                 & \multicolumn{1}{l|}{0.38}                                                                                   & 0.37                                                                                                \\ \cline{2-3} \cline{5-8} 
                                                                                     & \multicolumn{1}{l|}{0.008}       & 0                                &                                                                                                                                  & 12                                                                                                                                                 & 10                                                                                                                                 & \multicolumn{1}{l|}{0.29}                                                                                   & 0.27                                                                                                \\ \cline{2-3} \cline{5-8} 
                                                                                     & \multicolumn{1}{l|}{0.01}        & 0                                &                                                                                                                                  & 11                                                                                                                                                 & 8                                                                                                                                  & \multicolumn{1}{l|}{0.23}                                                                                   & 0.19                                                                                                \\ \cline{2-3} \cline{5-8} 
                                                                                     & \multicolumn{1}{l|}{0}           & 0.00002                          &                                                                                                                                  & 16                                                                                                                                                 & 15                                                                                                                                 & \multicolumn{1}{l|}{0.67}                                                                                   & 0.66                                                                                                \\ \cline{2-3} \cline{5-8} 
                                                                                     & \multicolumn{1}{l|}{0}           & 0.00004                          &                                                                                                                                  & 14                                                                                                                                                 & 13                                                                                                                                 & \multicolumn{1}{l|}{0.47}                                                                                   & 0.46                                                                                                \\ \cline{2-3} \cline{5-8} 
                                                                                     & \multicolumn{1}{l|}{0}           & 0.00006                          &                                                                                                                                  & 13                                                                                                                                                 & 11                                                                                                                                 & \multicolumn{1}{l|}{0.34}                                                                                   & 0.32                                                                                                \\ \cline{2-3} \cline{5-8} 
                                                                                     & \multicolumn{1}{l|}{0}           & 0.00008                          &                                                                                                                                  & 11                                                                                                                                                 & 9                                                                                                                                  & \multicolumn{1}{l|}{0.25}                                                                                   & 0.23                                                                                                \\ \cline{2-3} \cline{5-8} 
                                                                                     & \multicolumn{1}{l|}{0}           & 0.0001                           &                                                                                                                                  & 10                                                                                                                                                  & 8                                                                                                                                  & \multicolumn{1}{l|}{0.19}                                                                                   & 0.18                                                                                                \\ \cline{2-3} \cline{5-8} 
                                                                                     & \multicolumn{1}{l|}{0.002}       & 0.00002                          &                                                                                                                                  & 14                                                                                                                                                 & 14                                                                                                                                 & \multicolumn{1}{l|}{0.49}                                                                                   & 0.49                                                                                                \\ \cline{2-3} \cline{5-8} 
                                                                                     & \multicolumn{1}{l|}{0.004}       & 0.00004                          &                                                                                                                                  & 12                                                                                                                                                 & 11                                                                                                                                 & \multicolumn{1}{l|}{0.27}                                                                                   & 0.26                                                                                                \\ \cline{2-3} \cline{5-8} 
                                                                                     & \multicolumn{1}{l|}{0.006}       & 0.00006                          &                                                                                                                                  & 9                                                                                                                                                  & 8                                                                                                                                  & \multicolumn{1}{l|}{0.16}                                                                                   & 0.15                                                                                                \\ \cline{2-3} \cline{5-8} 
                                                                                     & \multicolumn{1}{l|}{0.008}       & 0.00008                          &                                                                                                                                  & 8                                                                                                                                                  & 5                                                                                                                                  & \multicolumn{1}{l|}{0.11}                                                                                   & 0.09                                                                                                \\ \cline{2-3} \cline{5-8} 
                                                                                     & \multicolumn{1}{l|}{0.01}        & 0.0001                           &                                                                                                                                  & 7                                                                                                                                                  & 4                                                                                                                                  & \multicolumn{1}{l|}{0.08}                                                                                   & 0.06                                                                                                \\ \hline
\end{tabular}
\end{adjustbox}
\end{table*}

We performed two sensitivity studies to understand the effect of noise on the probability of top outcome: (i) varying the type of noisy quantum gates, and (ii) varying the dimensionality of the problem or number of qubits. The results are shown in Figure~\ref{fig:gate_type_variation_result} and \ref{fig:num_qubit_variation_result} respectively. In each sub-figure, probability of the winning outcome is plotted against the number of iterations of amplitude amplification for three types of scenarios - (i) a completely noise-free simulation (shown by black solid line), (ii) the probability observed in noisy simulation by executing the transpiled circuit (shown by colored solid line), and (iii) the estimate probability of the winning outcome being noise free (shown by colored dashed line). Each subplot contains multiple lines corresponding to the simulated and estimated probabilities for varied noise parameter to understand - (i) the correlation between the number of qubits and the value of noise parameters, (ii) the correlation between the number of iterations and the value of noise parameters, and (iii) the correlation between the value of noise parameters and the outcome probability and in turn the optimal number of iteration where the peak probability is attained. 

\paragraphbe{Inflection point} It is rather apparent in both Figure~\ref{fig:gate_type_variation_result} and \ref{fig:num_qubit_variation_result} that there exists an inflection point iterating beyond which degrades the performance irrespective of - (i) the number of qubits to solve the problem, (ii) what type of quantum gates, noise is applied to, and (iii) the value of noise parameters. This reinforces our primary hypothesis motivating this work. However, the inflection point or the optimal number of iteration shifts further away from the theoretical optimal number of iterations in a noise-free environment with more number of qubits and higher value for noise paramater.

\paragraphbe{Strong correlation between simulation and estimation} We can also see that dashed, coloured lines closely resemble the solid, coloured lines. Note that the former corresponds to the collection of winning probabilities of being noise-free estimated by our proposed methods and the latter corresponds to the probability of correct outcome being observed from simulating a transpiled circuit. This strong correlation solidifies the usability of the estimated metric to predict the optimal number of iterations. However, there is always a gap between the solid and dashed coloured lines as they are measuring two different metrics. Note that in actual simulation the noise can result in some shots to be aligned with the winning output and this dynamic behavior is next to impossible to capture without an actual execution of the transpiled circuit. Hence, the probability of winning outcome obtained from a simulation is always slightly higher than the estimated probability of top-outcome being ``noise-free".

\paragraphbe{Accuracy of prediction} Table~\ref{tab:result} summarizes the experimental results by listing the optimal number of iterations observed in a noisy simulation and predicted by Method~\ref{alg:estimated_optimal} along with the probability of the winning outcome at both iterations reported by simulation for varied number of qubits and depolarizing noise parameters. We can see that the predicted value of optimal number of iterations is most of the times either same as the observed value or off by one. However, in some case with higher error rate, the predicted optimal number of iteration can be lower by up to three iterations from the optimal number of iterations reported by simulation, e.g., for problem composed of $9$ qubits and the \verb|sx| and \verb|cz| noise parameters respectively being $0.01$ and $0.0001$, the optimal number of iterations reported by simulation and predicted by Method~\ref{alg:estimated_optimal} are $7$ and $4$ respectively where the theoretical optimal number of iterations is $17$ in a noise-free environment. However, an interesting observation is that the outcome probability reported by simulation for iterations $7$ and $4$ are extremely close ($0.08\%$ and $0.06\%$ respectively). This observation is uniform throughout several data-points. It means that even when the user choose to simulated for the predicted number of iterations for amplitude amplification, there is not a significant loss from the peak attainable probability of the winning outcome.

\begin{table*}[h!t]
\caption{Optimal number of iterations predicted by analytical model, observed in noisy simulation, and on ibm\_oslo with their observed outcome probability.}
\label{tab:result_real_machine}
\begin{adjustbox}{width=\textwidth,keepaspectratio}
\begin{tabular}{|l|l|l|l|l|l|ll|ll|}
\hline
\multicolumn{1}{|c|}{\multirow{2}{*}{\textbf{Benchmark}}} & \multicolumn{1}{c|}{\multirow{2}{*}{\textbf{\# of qubits}}} & \multicolumn{1}{c|}{\multirow{2}{*}{\textbf{\begin{tabular}[c]{@{}c@{}}Theoretical\\ optimal \#\\ of iterations\end{tabular}}}} & \multicolumn{1}{c|}{\multirow{2}{*}{\textbf{\begin{tabular}[c]{@{}c@{}}Predicted\\ optimal \#\\ of iterations\end{tabular}}}} & \multicolumn{1}{c|}{\multirow{2}{*}{\textbf{\begin{tabular}[c]{@{}c@{}}Observed optimal \#\\ of iterations in\\ simulated env.\end{tabular}}}} & \multicolumn{1}{c|}{\multirow{2}{*}{\textbf{\begin{tabular}[c]{@{}c@{}}Observed optimal \#\\ of iterations on\\ ibm\_oslo\end{tabular}}}} & \multicolumn{2}{c|}{\textbf{\begin{tabular}[c]{@{}c@{}}Success probability\\ in simulated environment\end{tabular}}}                                                                                                                 & \multicolumn{2}{c|}{\textbf{\begin{tabular}[c]{@{}c@{}}Success probability\\ on ibm\_oslo\end{tabular}}}                                                                                                                             \\ \cline{7-10} 
\multicolumn{1}{|c|}{}                                    & \multicolumn{1}{c|}{}                                       & \multicolumn{1}{c|}{}                                                                                                        & \multicolumn{1}{c|}{}                                                                                                         & \multicolumn{1}{c|}{}                                                                                                                          & \multicolumn{1}{c|}{}                                                                                                                     & \multicolumn{1}{c|}{\textbf{\begin{tabular}[c]{@{}c@{}}For predicted\\ optimal \# \\ of iterations\end{tabular}}} & \multicolumn{1}{c|}{\textbf{\begin{tabular}[c]{@{}c@{}}For observed\\ optimal \# \\ of iterations\end{tabular}}} & \multicolumn{1}{c|}{\textbf{\begin{tabular}[c]{@{}c@{}}For predicted\\ optimal \# \\ of iterations\end{tabular}}} & \multicolumn{1}{c|}{\textbf{\begin{tabular}[c]{@{}c@{}}For observed\\ optimal \# \\ of iterations\end{tabular}}} \\ \hline
1-SAT                                                     & 5                                                           & 4                                                                                                                            & 2                                                                                                                             & 2                                                                                                                                              & 2                                                                                                                                         & \multicolumn{1}{l|}{0.222}                                                                                        & 0.222                                                                                                            & \multicolumn{1}{l|}{0.204}                                                                                        & 0.204                                                                                                            \\ \hline
3-SAT                                                     & 5                                                           & 2                                                                                                                            & 1                                                                                                                             & 1                                                                                                                                              & 1                                                                                                                                         & \multicolumn{1}{l|}{0.391}                                                                                        & 0.391                                                                                                            & \multicolumn{1}{l|}{0.363}                                                                                        & 0.363                                                                                                            \\ \hline
UOS                                                       & 5                                                           & 4                                                                                                                            & 2                                                                                                                             & 3                                                                                                                                              & 2                                                                                                                                         & \multicolumn{1}{l|}{0.252}                                                                                        & 0.257                                                                                                            & \multicolumn{1}{l|}{0.232}                                                                                        & 0.232                                                                                                            \\ \hline
\end{tabular}
\end{adjustbox}
\end{table*}

\begin{figure*}[h!t]
\begin{center}
\includegraphics[width=0.8\linewidth]{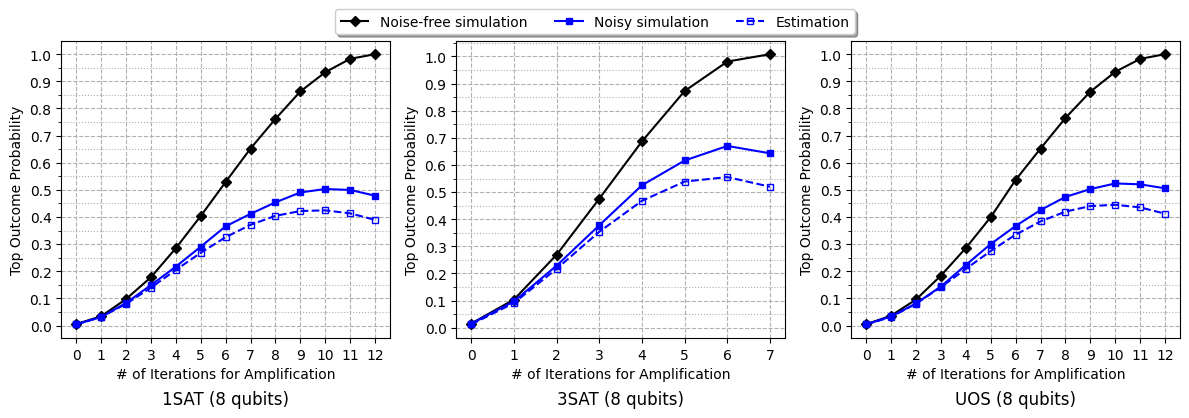}
\end{center}
\caption{Change in success probabilty with circuit repetition for more workloads with larger problem sizes on a quantum machine with single qubit and two-qubit noise rates as 0.004 and 0.00004 respectively.}
\label{fig:more_benchmark_result}
\end{figure*}

\paragraphbe{Correlation between the noise parameters, the number of qubits, and the number of gates} It is interesting to note that the depolarizing noise parameter single-qubit gate is varied from $0.002$ to $0.01$ in the increment of $0.002$ for experiments irrespective of the dimensionality of the problem. This is however not true for two-qubit quantum gates. The noise parameter for \verb|cx| gates ranges from $0.002$ to $0.01$ for $5$ qubit circuits, whereas they are two-order of magnitude smaller ranging from $2\times10^{-5}$ to $1\times10^{-4}$ with increment of $2\times10^{-5}$ for $9$-qubit circuits. This is because of two reasons. Firstly, as reported in Table~\ref{tab:gate_count}, the number of \verb|sx| gates is way smaller than the number of \verb|cx| gates in a transpiled circuit and the difference grows larger with higher number of qubits, e.g., for a circuit with $9$ qubit, the number of \verb|sx| gates is two-order of magnitude smaller than the number of \verb|cx| gates. As a result, loss of accuracy due to noise on \verb|sx| is relatively smaller than that from \verb|cx| gates. Second, the number of \verb|cx| gates grows by an order of magnitude from circuit with $5$ ($320$) to $7$ ($3008$) to $9$ ($25976$) qubits. So, to keep up with the growth in number of two-qubit gates with the increase in dimensionality of the problem, the noise parameter should also be scaled down by order of magnitude to yield any meaningful result and demonstrate the convex progression with an inflection point. If the value of noise-parameter is too high (e.g. noise parameters are $0.01$ and $0.0001$ for \verb|sx| and \verb|cx| gates in 9-qubit circuit) for the total number of quantum gates in the circuit, it results in an almost flat curve parallel to horizontal axis. This corresponds to all qubits being in a maximally mixed state where the
result is uniformly random irrespective of iterating over amplitude amplification.

\paragraphbe{Evaluation on real quantum machines}  We tried running our amplitude amplification programs on existing quantum machines on the IBM Quantum cloud.  We found that, even with our programs using just 5 qubits, all machines had error rates that were too high to provide any meaningful results.  Even after just one amplification iteration, the computation was overwhelmed by noise.  The noise was exacerbated by limited machine topologies where not all pairs of qubits allow entanglement, requiring many extra swap operations to be inserted to move qubits adjacent to each other.  So we decided to simulate a machine similar to ibm\_oslo with identical noise model including thermal relaxation as well as depolarizing noise, except with the modification that qubits are fully connected and allow arbitrary entanglement.  The results of this experiment is in Table~\ref{tab:result_real_machine}.  First, note that the ``Predicted optimal \# of iterations'' occurs earlier compared to the ``Theoretical optimal \# of iterations'', in line with previous results.  Importantly, note that the prediction is not only accurate for the ``simulated environment", but also for ibm\_oslo with the full noise model, the only exception being UOS (Un-Ordered Search) with a difference of 1 iteration.  And even for UOS, the difference between the success probability when using our prediction (0.252) and on the actual observed optimal point (0.257) is negligible. This shows the efficacy of predicting the optimal number of iterations by our analytical model even when noise channels other than depolarizing quantum noise is present.  We plan to eventually add thermal relaxation to our prediction model in future work, but we feel these are promising results.

\paragraphbe{Evaluation on other benchmarks}  In Figure~\ref{fig:more_benchmark_result}, we show the results of applying our optimization to 3-SAT and UOS (Un-Ordered Search) along with 1-SAT, using a program size of 8 qubits on a machine with single qubit and two-qubit noise rates as $0.004$ and $0.00004$ respectively.  Note that, just like previous results, the observed simulation curve closely tracks our estimated prediction curve.  The theoretical optimal iterations of 1-SAT, 3-SAT and UOS are 12, 7, and 12 respectively.  The predicted optimal iterations (the peak of the dotted lines) are 10, 6, and 10 respectively which match exactly with the observed optimal iterations (the peak of the solid lines).  This shows that our optimization is applicable to a variety of quantum algorithms that inherently repeats part of circuit with more iterations.
\section{Related Work}
\label{sec:related_work}

The first body of work studies the relationship between amplitude amplification algorithms and various types of noise~\cite{cohn2016grovers,reitzner2019Grover,PabloNorman1999Noise,rastegin2017Degradation,kravchenko2016Grovers}. Work by Ilan et al.~\cite{cohn2016grovers} on total depolarizing and local depolarizing channels is perhaps the closest related work.  It showed that total depolarizing noise rate smaller than $\frac{1}{\sqrt{N}}$ and a local depolarizing error rate smaller than $\frac{1}{\sqrt{N}\log_2 N}$ allows Grover's algorithm to run without error correction.  Work by Reitzner et al.~\cite{reitzner2019Grover} studies the effect of localized dephasing noise of rate $p$ on an affected subspace of size $k$ on Grover's algorithm. When the searched target is unaffected by noise, $kp << \sqrt{N}$ is sufficient to achieve quadratic speedup. When the target is affected, the noise rate needs to scale with $\frac{1}{\sqrt{N}}$.  Work by Pablo-Norman et al.~\cite{PabloNorman1999Noise} studies the effect of random Gaussian noise on Grover's algorithm and concludes that a noise rate of $N^{-2/3}$ is sufficient to achieve speedup.

The second body of work encompasses efforts to optimize Grover's algorithm (and by extension other amplitude amplification algorithms) to better withstand noise~\cite{Wang2020Prospect, Zhang2020Depth}.  These studies focus on reducing the ``depth'', or the number of gates, needed to compose the circuit using various techniques.  The work by Zhang et al.~\cite{Zhang2020Depth} employs a divide-and-conquer approach where the search algorithm is divided into several stages where each stage is shallower in depth than the original circuit.  A ``local'' diffusion operator, that applies to only a subspace of the qubits, is used in lieu of the global diffusion operator used by conventional Grover's algorithm.  The work by Wang et al.~\cite{Wang2020Prospect} reduces the circuit depth by the use of multi-control Toffolis modified by addition of an ancilla qubit.  The work shows that further improvement is achievable beyond that possible by local diffusion operators for various types of quantum errors.

The first body of work is theoretical and are limit-studies on how various noise types must scale with N in order to achieve the quadratic speedup offered by amplitude amplification without error correction.  To an extent, our empirical results confirm the theoretical results on the existence of an inflection point in the presence of noise~\cite{cohn2016grovers}.  But to the best of our knowledge, there is no previous work that deals with the practical question: ``Given a quantum algorithm and a quantum machine, what is the best accuracy the user can expect from the machine?".  We propose a concrete framework using novel transpiler intrinsics that answers that question with high accuracy and also optimizes the circuit to achieve that accuracy.  Our optimization can be used in conjunction with previous optimizations cited in the second body of work.  As long as the iterative structure of the algorithm does not change, it can benefit from our optimization to find the iteration at the inflection point.
\section{Conclusion}
\label{sec:conclusion}

Through detailed empirical observation, we successfully demonstrated that with every iteration of amplitude amplification, the probability of the winning outcome increases  until it reaches an ``inflection" point beyond which the degrading effects of noise surmounts the benefit from amplitude amplification. We implemented a noise-aware extension to the transpiler that takes the quantum depolarizing noise parameters for each quantum gate into consideration to determine the optimal number of iterations for amplitude amplification at which the winning outcome attains the peak probability.  We do this by calculating the probability of noise-free outcome at each iteration, as a proxy for the probability of accurate winning outcome.  We empirically show that the former closely tracks the latter for different sizes of the Boolean satisfiability problem (1-SAT), making our prediction of the inflection point also accurate.

We would like to stress that this prediction is made by pure analysis proportional in time to the number of gates in the circuit, and there is no need to do any type of ``profiling'' by executing the circuit on either a quantum simulator or a real quantum machine. It is to be noted that quantum machines advertise both depolarizing gate noise and thermal relaxation error. This work only models the effect of depolarizing gate noise on amplitude amplification. In the future, we would like to investigate the effect of thermal relaxation and integrate it into the prediction framework.

\normalsize
\bibliographystyle{IEEEtranS}
\bibliography{refs}

\begin{thebibliography}{10}
\providecommand{\url}[1]{#1}
\csname url@samestyle\endcsname
\providecommand{\newblock}{\relax}
\providecommand{\bibinfo}[2]{#2}
\providecommand{\BIBentrySTDinterwordspacing}{\spaceskip=0pt\relax}
\providecommand{\BIBentryALTinterwordstretchfactor}{4}
\providecommand{\BIBentryALTinterwordspacing}{\spaceskip=\fontdimen2\font plus
\BIBentryALTinterwordstretchfactor\fontdimen3\font minus
  \fontdimen4\font\relax}
\providecommand{\BIBforeignlanguage}[2]{{%
\expandafter\ifx\csname l@#1\endcsname\relax
\typeout{** WARNING: IEEEtranS.bst: No hyphenation pattern has been}%
\typeout{** loaded for the language `#1'. Using the pattern for}%
\typeout{** the default language instead.}%
\else
\language=\csname l@#1\endcsname
\fi
#2}}
\providecommand{\BIBdecl}{\relax}
\BIBdecl

\bibitem{IBMQiskit}
M.~S. ANIS, Abby-Mitchell, H.~Abraham, AduOffei, R.~Agarwal, G.~Agliardi,
  M.~Aharoni, V.~Ajith, I.~Y. Akhalwaya, G.~Aleksandrowicz, T.~Alexander,
  M.~Amy, S.~Anagolum, Anthony-Gandon, E.~Arbel, A.~Asfaw, A.~Athalye,
  A.~Avkhadiev, C.~Azaustre, P.~BHOLE, A.~Banerjee, S.~Banerjee, W.~Bang,
  A.~Bansal, P.~Barkoutsos, A.~Barnawal, G.~Barron, G.~S. Barron, L.~Bello,
  Y.~Ben-Haim, M.~C. Bennett, D.~Bevenius, D.~Bhatnagar, P.~Bhatnagar,
  A.~Bhobe, P.~Bianchini, L.~S. Bishop, C.~Blank, S.~Bolos, S.~Bopardikar,
  S.~Bosch, S.~Brandhofer, Brandon, S.~Bravyi, N.~Bronn, Bryce-Fuller,
  D.~Bucher, A.~Burov, F.~Cabrera, P.~Calpin, L.~Capelluto, J.~Carballo,
  G.~Carrascal, A.~Carriker, I.~Carvalho, A.~Chen, C.-F. Chen, E.~Chen, J.~C.
  Chen, R.~Chen, F.~Chevallier, K.~Chinda, R.~Cholarajan, J.~M. Chow,
  S.~Churchill, CisterMoke, C.~Claus, C.~Clauss, C.~Clothier, R.~Cocking,
  R.~Cocuzzo, J.~Connor, F.~Correa, Z.~Crockett, A.~J. Cross, A.~W. Cross,
  S.~Cross, J.~Cruz-Benito, C.~Culver, A.~D. C{\'o}rcoles-Gonzales, N.~D,
  S.~Dague, T.~E. Dandachi, A.~N. Dangwal, J.~Daniel, M.~Daniels, M.~Dartiailh,
  A.~R. Davila, F.~Debouni, A.~Dekusar, A.~Deshmukh, M.~Deshpande, D.~Ding,
  J.~Doi, E.~M. Dow, P.~Downing, E.~Drechsler, E.~Dumitrescu, K.~Dumon,
  I.~Duran, K.~EL-Safty, E.~Eastman, G.~Eberle, A.~Ebrahimi, P.~Eendebak,
  D.~Egger, ElePT, Emilio, A.~Espiricueta, M.~Everitt, D.~Facoetti, Farida,
  P.~M. Fern{\'a}ndez, S.~Ferracin, D.~Ferrari, A.~H. Ferrera, R.~Fouilland,
  A.~Frisch, A.~Fuhrer, B.~Fuller, M.~GEORGE, J.~Gacon, B.~G. Gago,
  C.~Gambella, J.~M. Gambetta, A.~Gammanpila, L.~Garcia, T.~Garg, S.~Garion,
  J.~R. Garrison, J.~Garrison, T.~Gates, H.~Georgiev, L.~Gil, A.~Gilliam,
  A.~Giridharan, Glen, J.~Gomez-Mosquera, Gonzalo, S.~de~la
  Puente~Gonz{\'a}lez, J.~Gorzinski, I.~Gould, D.~Greenberg, D.~Grinko,
  W.~Guan, D.~Guijo, Guillermo-Mijares-Vilarino, J.~A. Gunnels, H.~Gupta,
  N.~Gupta, J.~M. G{\"u}nther, M.~Haglund, I.~Haide, I.~Hamamura, O.~C. Hamido,
  F.~Harkins, K.~Hartman, A.~Hasan, V.~Havlicek, J.~Hellmers, {\L}.~Herok,
  S.~Hillmich, C.~Hong, H.~Horii, C.~Howington, S.~Hu, W.~Hu, C.-H. Huang,
  J.~Huang, R.~Huisman, H.~Imai, T.~Imamichi, K.~Ishizaki, Ishwor, R.~Iten,
  T.~Itoko, A.~Ivrii, A.~Javadi, A.~Javadi-Abhari, W.~Javed, Q.~Jianhua,
  M.~Jivrajani, K.~Johns, S.~Johnstun, Jonathan-Shoemaker, JosDenmark,
  JoshDumo, J.~Judge, T.~Kachmann, A.~Kale, N.~Kanazawa, J.~Kane, Kang-Bae,
  A.~Kapila, A.~Karazeev, P.~Kassebaum, T.~Kehrer, J.~Kelso, S.~Kelso, H.~van
  Kemenade, V.~Khanderao, S.~King, Y.~Kobayashi, Kovi11Day, A.~Kovyrshin,
  R.~Krishnakumar, P.~Krishnamurthy, V.~Krishnan, K.~Krsulich, P.~Kumkar,
  G.~Kus, R.~LaRose, E.~Lacal, R.~Lambert, H.~Landa, J.~Lapeyre, J.~Latone,
  S.~Lawrence, C.~Lee, G.~Li, T.~J. Liang, J.~Lishman, D.~Liu, P.~Liu, Lolcroc,
  A.~K. M, L.~Madden, Y.~Maeng, S.~Maheshkar, K.~Majmudar, A.~Malyshev, M.~E.
  Mandouh, J.~Manela, Manjula, J.~Marecek, M.~Marques, K.~Marwaha, D.~Maslov,
  P.~Maszota, D.~Mathews, A.~Matsuo, F.~Mazhandu, D.~McClure, M.~McElaney,
  C.~McGarry, D.~McKay, D.~McPherson, S.~Meesala, D.~Meirom, C.~Mendell,
  T.~Metcalfe, M.~Mevissen, A.~Meyer, A.~Mezzacapo, R.~Midha, D.~Miller,
  H.~Miller, Z.~Minev, A.~Mitchell, N.~Moll, A.~Montanez, G.~Monteiro, M.~D.
  Mooring, R.~Morales, N.~Moran, D.~Morcuende, S.~Mostafa, M.~Motta, R.~Moyard,
  P.~Murali, D.~Murata, J.~M{\"u}ggenburg, T.~NEMOZ, D.~Nadlinger,
  K.~Nakanishi, G.~Nannicini, P.~Nation, E.~Navarro, Y.~Naveh, S.~W. Neagle,
  P.~Neuweiler, A.~Ngoueya, T.~Nguyen, J.~Nicander, Nick-Singstock, P.~Niroula,
  H.~Norlen, NuoWenLei, L.~J. O'Riordan, O.~Ogunbayo, P.~Ollitrault,
  T.~Onodera, R.~Otaolea, S.~Oud, D.~Padilha, H.~Paik, S.~Pal, Y.~Pang,
  A.~Panigrahi, V.~R. Pascuzzi, S.~Perriello, E.~Peterson, A.~Phan, K.~Pilch,
  F.~Piro, M.~Pistoia, C.~Piveteau, J.~Plewa, P.~Pocreau, A.~Pozas-Kerstjens,
  R.~Pracht, M.~Prokop, V.~Prutyanov, S.~Puri, D.~Puzzuoli, Pythonix,
  J.~P{\'e}rez, Quant02, Quintiii, R.~I. Rahman, A.~Raja, R.~Rajeev, I.~Rajput,
  N.~Ramagiri, A.~Rao, R.~Raymond, O.~Reardon-Smith, R.~M.-C. Redondo,
  M.~Reuter, J.~Rice, M.~Riedemann, Rietesh, D.~Risinger, P.~Rivero, M.~L.
  Rocca, D.~M. Rodr{\'\i}guez, RohithKarur, B.~Rosand, M.~Rossmannek, M.~Ryu,
  T.~SAPV, N.~R.~C. Sa, A.~Saha, A.~Ash-Saki, S.~Sanand, M.~Sandberg,
  H.~Sandesara, R.~Sapra, H.~Sargsyan, A.~Sarkar, N.~Sathaye, N.~Savola,
  B.~Schmitt, C.~Schnabel, Z.~Schoenfeld, T.~L. Scholten, E.~Schoute,
  M.~Schulterbrandt, J.~Schwarm, J.~Seaward, Sergi, I.~F. Sertage, K.~Setia,
  F.~Shah, N.~Shammah, W.~Shanks, R.~Sharma, P.~Shaw, Y.~Shi, J.~Shoemaker,
  A.~Silva, A.~Simonetto, D.~Singh, D.~Singh, P.~Singh, P.~Singkanipa,
  Y.~Siraichi, Siri, J.~Sistos, I.~Sitdikov, S.~Sivarajah, Slavikmew, M.~B.
  Sletfjerding, J.~A. Smolin, M.~Soeken, I.~O. Sokolov, I.~Sokolov, V.~P.
  Soloviev, SooluThomas, Starfish, D.~Steenken, M.~Stypulkoski, A.~Suau,
  S.~Sun, K.~J. Sung, M.~Suwama, O.~S{\l}owik, H.~Takahashi, T.~Takawale,
  I.~Tavernelli, C.~Taylor, P.~Taylour, S.~Thomas, K.~Tian, M.~Tillet, M.~Tod,
  M.~Tomasik, C.~Tornow, E.~de~la Torre, J.~L.~S. Toural, K.~Trabing,
  M.~Treinish, D.~Trenev, TrishaPe, F.~Truger, G.~Tsilimigkounakis, D.~Tulsi,
  D.~Tuna, W.~Turner, Y.~Vaknin, C.~R. Valcarce, F.~Varchon, A.~Vartak, A.~C.
  Vazquez, P.~Vijaywargiya, V.~Villar, B.~Vishnu, D.~Vogt-Lee, C.~Vuillot,
  J.~Weaver, J.~Weidenfeller, R.~Wieczorek, J.~A. Wildstrom, J.~Wilson,
  E.~Winston, WinterSoldier, J.~J. Woehr, S.~Woerner, R.~Woo, C.~J. Wood,
  R.~Wood, S.~Wood, J.~Wootton, M.~Wright, L.~Xing, J.~YU, Yaiza, B.~Yang,
  U.~Yang, J.~Yao, D.~Yeralin, R.~Yonekura, D.~Yonge-Mallo, R.~Yoshida,
  R.~Young, J.~Yu, L.~Yu, Yuma-Nakamura, C.~Zachow, L.~Zdanski, H.~Zhang,
  I.~Zidaru, B.~Zimmermann, C.~Zoufal, aeddins ibm, alexzhang13, b63, bartek
  bartlomiej, bcamorrison, brandhsn, chetmurthy, deeplokhande, dekel.meirom,
  dime10, dlasecki, ehchen, ewinston, fanizzamarco, fs1132429, gadial,
  galeinston, georgezhou20, georgios ts, gruu, hhorii, hhyap, hykavitha, itoko,
  jeppevinkel, jessica angel7, jezerjojo14, jliu45, johannesgreiner, jscott2,
  kUmezawa, klinvill, krutik2966, ma5x, michelle4654, msuwama, nico lgrs,
  nrhawkins, ntgiwsvp, ordmoj, sagar pahwa, pritamsinha2304, rithikaadiga,
  ryancocuzzo, saktar unr, saswati qiskit, septembrr, sethmerkel, sg495,
  shaashwat, smturro2, sternparky, strickroman, tigerjack, tsura crisaldo,
  upsideon, vadebayo49, welien, willhbang, wmurphy collabstar, yang.luh, and
  M.~{\v{C}}epulkovskis, ``Qiskit: An open-source framework for quantum
  computing,'' 2021.

\bibitem{bera2018amplitude}
D.~Bera, ``Amplitude amplification for operator identification and randomized
  classes,'' in \emph{International Computing and Combinatorics
  Conference}.\hskip 1em plus 0.5em minus 0.4em\relax Springer, 2018, pp.
  579--591.

\bibitem{brassard1997exact}
G.~Brassard and P.~Hoyer, ``An exact quantum polynomial-time algorithm for
  simon's problem,'' in \emph{Proceedings of the Fifth Israeli Symposium on
  Theory of Computing and Systems}.\hskip 1em plus 0.5em minus 0.4em\relax
  IEEE, 1997, pp. 12--23.

\bibitem{brassard2002quantum}
G.~Brassard, P.~Hoyer, M.~Mosca, and A.~Tapp, ``Quantum amplitude amplification
  and estimation,'' \emph{Contemporary Mathematics}, vol. 305, pp. 53--74,
  2002.

\bibitem{cohn2016grovers}
I.~Cohn, A.~Fonseca~de Oliveira, E.~Buksman, and J.~Lacalle, ``Grover's search
  with local and total depolarizing channel errors: Complexity analysis,''
  \emph{International Journal of Quantum Information}, vol.~14, no.~2, p.
  1650009, 2016.

\bibitem{daoyi2007reinforcement}
D.~Daoyi, C.~Chunlin, and L.~Hanxiong, ``Reinforcement strategy using quantum
  amplitude amplification for robot learning,'' in \emph{2007 Chinese Control
  Conference}.\hskip 1em plus 0.5em minus 0.4em\relax IEEE, 2007, pp. 571--575.

\bibitem{elias2021enhanced}
B.~Elias and A.~Younes, ``Enhanced quantum signature scheme using quantum
  amplitude amplification operators,'' \emph{PloS one}, vol.~16, no.~10, p.
  e0258091, 2021.

\bibitem{franco2009quantum}
R.~Franco, ``Quantum amplitude amplification algorithm: an explanation of
  availability bias,'' in \emph{International Symposium on Quantum
  Interaction}.\hskip 1em plus 0.5em minus 0.4em\relax Springer, 2009, pp.
  84--96.

\bibitem{grover1996fast}
L.~K. Grover, ``A fast quantum mechanical algorithm for database search,'' in
  \emph{Proceedings of the twenty-eighth annual ACM symposium on Theory of
  computing}, 1996, pp. 212--219.

\bibitem{hastings2009superadditivity}
M.~B. Hastings, ``Superadditivity of communication capacity using entangled
  inputs,'' \emph{Nature Physics}, vol.~5, no.~4, pp. 255--257, 2009.

\bibitem{IBMQResources}
{IBM Q team}, ``{IBM Q backend specification},''
  \url{https://quantum-computing.ibm.com}.

\bibitem{king2003capacity}
C.~King, ``The capacity of the quantum depolarizing channel,'' \emph{IEEE
  Transactions on Information Theory}, vol.~49, no.~1, pp. 221--229, 2003.

\bibitem{koch2022gaussian}
D.~Koch, M.~Cutugno, S.~Karlson, S.~Patel, L.~Wessing, and P.~M. Alsing,
  ``Gaussian amplitude amplification for quantum pathfinding,'' \emph{Entropy},
  vol.~24, no.~7, p. 963, 2022.

\bibitem{kravchenko2016Grovers}
D.~Kravchenko, N.~Nahimovs, and A.~Rivosh, ``Grover's search with faults on
  some marked elements,'' in \emph{International Conference on SOFSEM 2016:
  Theory and Practice of Computer Science}.\hskip 1em plus 0.5em minus
  0.4em\relax ACM, 2016.

\bibitem{kwon2021quantum}
H.~Kwon and J.~Bae, ``Quantum amplitude-amplification operators,''
  \emph{Physical Review A}, vol. 104, no.~6, p. 062438, 2021.

\bibitem{mckay2017efficient}
D.~C. McKay, C.~J. Wood, S.~Sheldon, J.~M. Chow, and J.~M. Gambetta,
  ``Efficient z gates for quantum computing,'' \emph{Physical Review A},
  vol.~96, no.~2, p. 022330, 2017.

\bibitem{PabloNorman1999Noise}
B.~Pablo-Norman and M.~Ruiz-Altaba, ``Noise in grover's quantum search
  algorithm,'' \emph{Physical Review A}, vol.~61, no.~1, 1999.

\bibitem{rajagopal2021quantum}
K.~Rajagopal, Q.~Zhang, S.~Balakrishnan, P.~Fakhari, and J.~Busemeyer,
  ``Quantum amplitude amplification for reinforcement learning,''
  \emph{Handbook of Reinforcement Learning and Control}, pp. 819--833, 2021.

\bibitem{rastegin2017Degradation}
A.~Rastegin, ``On degradation of grover's search under collective phase flips
  in queries to the oracle,'' \emph{Frontiers of Physics}, vol.~13, 2017.

\bibitem{reitzner2019Grover}
D.~Reitzner and M.~Hillery, ``Grover search under localized dephasing,''
  \emph{Physical Review A}, vol.~99, p. 012339, 2019.

\bibitem{Wang2020Prospect}
Y.~Wang and P.~S. Krstic, ``Prospect of using grover's search in the
  noisy-intermediate-scale quantum-computer era,'' \emph{Physical Review A},
  vol. 102, p. 042609, 2020.

\bibitem{Zhang2020Depth}
K.~Zhang and V.~E. Korepin, ``Depth optimization of quantum search algorithms
  beyond grover's algorithm,'' \emph{Phys. Rev. A}, vol. 101, p. 032346, 2020.

\end{thebibliography}

\end{document}